%% file: main.tex
\documentclass[twocolumn]{aastex63}

\input{./preamble.tex}

\newcommand{\pa}{^{\prime}}

\submitjournal{ApJ}

\shorttitle{Polydisperse Pebble Accretion}
\shortauthors{Lyra et al.}

\begin{document}

\title{An Analytical Theory for the Growth from Planetesimals to
  Planets by Polydisperse Pebble Accretion}

\correspondingauthor{Wladimir Lyra}
\email{wlyra@nmsu.edu}

\author[0000-0002-3768-7542]{Wladimir Lyra}
\affiliation{New Mexico State University, Department of Astronomy, PO Box 30001 MSC 4500, Las Cruces, NM 88001, USA}

\author[0000-0002-5893-6165]{Anders Johansen}
\affiliation{Center for Star and Planet Formation, GLOBE Institute,
  University of Copenhagen, Øster Voldgade 5-7, 1350 Copenhagen,
  Denmark}
\affiliation{Lund Observatory, Department of Astronomy and Theoretical Physics, Lund University, Box 43, 221 00 Lund, Sweden}

\author{Manuel H. Ca\~nas}
\affiliation{New Mexico State University, Department of Astronomy, PO Box 30001 MSC 4500, Las Cruces, NM 88001, USA}

%\author[0000-0003-2589-5034]{Chao-Chin Yang (\cntext{楊朝欽})}
\author[0000-0003-2589-5034]{Chao-Chin Yang}
\affiliation{Department of Physics and Astronomy, University of Alabama, Box 870324, Tuscaloosa, AL 35487-0324, USA}

\begin{abstract}

Pebble accretion is recognized as a significant accelerator of planet
formation. Yet, only formulae for single-sized (monodisperse)
distribution have been derived in the literature. These can lead to
significant underestimates for Bondi accretion, for which the best
accreted pebble size may not be the one that dominates the mass
distribution. We derive in this paper the polydisperse theory of
pebble accretion. We consider a power-law distribution in pebble
  radius, and we find the resulting surface and volume number density distribution
functions. We derive also the exact monodisperse analytical pebble
accretion rate for which 3D and 2D accretion are
  limits. In addition, we find analytical solutions to the polydisperse 2D Hill and 3D Bondi
limits. We integrate the polydisperse
pebble accretion numerically for the MRN distribution, finding a slight decrease (by an exact
factor 3/7) in the Hill regime compared to the monodisperse case. In contrast, in the Bondi regime, we
find 1-2 orders of magnitude higher
accretion rates compared to monodisperse, also
extending the onset of pebble accretion to 1-2 order of magnitude
lower in mass. We find Myr-timescales, within the disk lifetime, for
Bondi accretion on top of planetary seeds of masses
$10^{-6}-10^{-4} M_\oplus$, over a significant range of the parameter
space. This mass range overlaps with the high mass end of the
planetesimal initial mass function, and thus pebble accretion is
possible directly following formation by streaming instability. This alleviates the need for mutual planetesimal collisions as a major contribution to planetary growth.

\end{abstract}


\section{Introduction}

Despite significant theoretical and observational advances in the past
decade, a comprehensive theory of planet formation still remains
elusive. Planet formation starts from the accumulation of sub-$\mu$m
interstellar grains, growing by means of coagulation, in hit-and-stick
low-velocity collisions
\citep{Safronov72,Nakagawa+81,Tominaga+21}. Laboratory experiments
\citep{BlumWurm08,Guttler+10} and numerical simulations
\citep{Guttler+09,Geretshauser+10,Zsom+10} provide evidence that this process
is efficient in growing solid grains up to mm and cm radius (hereafter called
``pebbles'') with growth beyond this size being unlikely, due to bouncing,
fragmentation, and drift
\citep{DullemondDominik05,Brauer+08,Krijt+15}, unless the
  possibility of very high porosities is introduced \citep{Suyama+08,Suyama+12}.

The streaming instability \citep{YoudinGoodman05,YoudinJohansen07, JohansenYoudin07,Kowalik+13,LyraKuchner13,Krapp+19,SquireHopkins20,Schaefer+20,Paardekooper+20,ChenLin20,McNally+21,Lin21,FlockMignone21,ZhuYang21,YangZhu21} whereby the drift of grains through the gas
is unstable, has been established as a mechanism to produce the first
planetesimals
\citep{Johansen+07,YangJohansen14,Carrera+15,Simon+16,Yang+17,Schaffer+18,Nesvorny+19,Li+19,KlahrSchreiber21,Visser+21,LiYoudin21},
through concentration of pebbles into dense
  filaments that display a fractal structure with large overdensities
  reached at the smallest scales of the simulations
  \citep{Johansen+15}. Yet, growth by binary accretion of planetesimals
into progressively larger objects, while able to explain the growth of
a giant planet's core at 5\,AU \citep[if migration is ignored,][]{Pollack+94}, is not viable
already at the orbital position
of Saturn, Uranus, or Neptune \citep{Thommes+03,JohansenBitsch19}.

This shortcoming of planetesimal accretion motivated the search for other
avenues of planetary growth. Fast accretion rates of marginally coupled
solids up to planetary masses were first seen in the simulations of
\citet{Lyra+08b}.  In that model, vortices trap pebbles and
collapse them into Moon-mass objects via direct gravitational
instability, which scoop up the remaining pebbles at a vertiginous
rate, achieving Mars and Earth mass within a few hundred orbits. Whereas this growth was assisted by vortices, it
illustrates that gas-assisted accretion of pebbles is potentially much faster
than planetesimal accretion, due to the presence of gas drag as a dissipative
mechanism. A similar result was found by \cite{JohansenLacerda10},
showing fast accretion rates onto a 100\,km seed, highlighting the
importance of pebble accretion for planetary growth, and suggesting for the first time that
a significant fraction of the accretion of planetary bodies proceeds via
pebbles (as opposed to planetesimals), before the dissipation of the
gas disk.

An analytical theory of pebble accretion was later developed by
\citet{OrmelKlahr10} and \citet{LambrechtsJohansen12}, elucidating the
existence of two regimes: one for small masses, where the seed mass accretes from a 
pebble headwind, a process reminiscent of Bondi-Hoyle-Lyttleton accretion \citep{Bondi44,HoyleLyttleton39}; and
another, for higher masses, where pebbles are accreted from the whole
Hill sphere of the seed. These regimes were dubbed
  ``drift-dominated'' and ``shear-dominated'' by \citet{OrmelKlahr10},
  respectively, whereas \cite{LambrechtsJohansen12} called them
  ``Bondi'' and ``Hill''. As a rule of thumb, planetesimals accrete in the Bondi regime, 
protoplanets in the Hill regime \citep{Ormel17,JohansenLambrechts17},
and both can yield orders-of-magnitude higher mass accretion rates than planetesimal
accretion.

Since its inception, the model has quickly risen to paradigmatic status, by
virtue of a number of successes. Pebble accretion explains the
formation of the gas giants \citep{LambrechtsJohansen12},
    of the ice giants with low gas fractions \citep{Lambrechts+14}; the preponderance of
super-Earths around other stars \citep{Lambrechts+19,Bitsch+19b,Izidoro+21}; it achieves a
better planet population synthesis matching exoplanet
populations than a planetesimal-based accretion model
\citep{Bitsch+19,Drazkowska+22}, and it is
also compatible with the drift-dominated evolution of dust in T-Tauri
disks (a flux of $\sim$ 100 Earth masses over the disk lifetime, 
\citealt{Appelgren+20}). Even the classical giant impact model for
  terrestrial planet formation \citep{Raymond+04}
  is challenged now by a hybrid view where terrestrial planets
  accrete their mass from a combination of planetesimals and small
  pebbles \citep{Johansen+15,Johansen+21}. 

However, most previous works on pebble accretion considered a monodisperse
distribution of pebbles. In reality, the pebbles will have a
distribution of sizes, ranging from sub-$\mu$m to mm or cm-size.
A monodisperse distribution can be a reasonable assumption because, for
the interstellar grain size distribution, following a power-law of -3.5 of the
grain radius \citep[][MRN henceforth]{Mathis+77,HirashitaKobayashi13} most of the
mass resides in the largest pebbles; a result that stands even after dust
evolution away from MRN in the protoplanetary disk is considered
\citep{Birnstiel+12a}. This makes the Hill regime of pebble accretion relatively
insensitive to the dust spectrum, and either the dominant pebble
size \citep{LambrechtsJohansen14} or a mass weighted representative
pebble size \citep{Guilera+20,Venturini+20} yield sensible
results.

Indeed, in a recent work, \citet{Andama+22}, considering
polydisperse Hill accretion, find larger final core masses, not
because of faster accretion rates, but because the smaller grains
drift more slowly, lingering around for longer times than the largest
pebbles, and thus extending the duration of
accretion. \citet{Drazkowska+21} also considering the Hill regime, focus on
the beneficial aspects of fragmentation on keeping the pebbles
sizes small, because too large pebbles accrete poorly. Both
works consider a body already near the Bondi-Hill transition mass, a polydisperse size spectrum
from the prescription of \citet{Birnstiel+12a}, and solve numerically for
the mass accretion rates. Both works also highlight how the mass accretion rate is
dependent on the embryo mass but not on pebble size.

In stark constrast, in the Bondi regime the size distribution
should matter significantly for the mass accretion rate itself. In the Bondi regime, the best accreted pebbles
are those of friction time similar to the time the pebble takes to
cross the Bondi radius, i.e., the Bondi time. For small enough
seed mass, the larger, cm-sized, pebbles, drift so fast past the protoplanet
that these pebbles essentially behave like planetesimals. In this case, the
cross section for accretion is geometric (for high speeds), or
  gravitationally focused (for low speeds), and only slightly aided by
  gas drag. As a result, even though these pebbles
dominate the mass budget, their mass accretion rate by the planetesimal can be lower
than of the smaller pebbles for which Bondi accretion is more
efficient. If that is the case, the pebble accretion rates in the Bondi
regime may be underestimated by the current monodisperse
prescriptions. Indeed, \citet{LorekJohansen22} recently find
  that planetesimal accretion is insignificant beyond 5\,AU, so the
  onset of pebble accretion has to overlap with the high-mass end of the planetesimal
  mass function if planet formation is to proceed.

In this paper, we work out the polydisperse extension of
pebble accretion. We find that indeed Bondi accretion is 1-2
orders of magnitude more efficient in the polydisperse
case. We also find that the onset of polydisperse Bondi accretion occurs at
  lower masses than monodisperse, by 1-2 orders of magnitude. Hill
accretion is slightly less efficient, by a factor 3/7, for the
  MRN distribution. We find
  the exact solution to the 2D-3D transition, as well as analytical
  expressions for the polydisperse 2D Hill and 3D Bondi accretion
  rates.

This paper is structured as follows. In \sect{sect:model} we derive
the grain size distribution functions; in \sect{sect:pebbleaccretion}
we apply them to pebble accretion, deriving the polydisperse
model, and proceeding with the analysis. In
  \sect{sect:analytical} we work out the analytical expressions for
  2D Hill and 3D Bondi polydisperse accretion. A summary concludes the
  paper in \sect{sect:conclusion}. A table of mathematical symbols used in this work is shown in \table{table:symbols}.

   \begin{table*}
\caption{Symbols used in this work.}
\label{table:symbols}
\begin{center}
\begin{tabular}{lll c lll}\hline
  Symbol & Definition & Description & &  Symbol & Definition & Description \\\hline
  $F$&\eq{eq:BigF}&pebble size distribution                             && $\rho_{d0}$&\eq{eq:rhod0-midplane}&dust density at midplane\\
  $a$&&pebble radius                                                    && $\delta v$&\eq{eq:deltav}&approach velocity\\
  $z$&&vertical coordinate                                              && $S$&\eq{eq:stratification-analytical}&stratification integral\\
  $n$&\eq{eq:numberdensity} &number density                             && $\Delta v$&&Sub-Keplerian velocity reduction\\
  $m$&\eq{eq:pebblemass}&pebble mass                                    && $\varOmega$&$\sqrt{\frac{GM_\odot}{r^3}}$&Keplerian frequency\\        
  $\rho_d$&\eq{eq:rhoz} &volume density                                 && $\hat{R}_{\rm acc}$&\eq{eq:racchatgeneral}&accretion radius\\
  $H_d$&\eq{eq:Hd}&pebble scale height                                  && $\chi$&\eq{eq:racc}&coefficient\\
  $f$&\eq{eq:fada} &pebble distribution in midplane                     && $\tau_f$&$\St/\varOmega$&friction time\\
  $H_g$&$\varOmega/c_s$&gas scale height                                && $t_p$&\eq{eq:tp}&passing timescale\\             
  $\alpha$&&Shakura-Sunyaev viscosity                                   && $\gamma$&\eq{eq:racc}&coefficient\\
  $\St$&\eq{eq:stokes}&Stoker number                                    && $G$&&gravitational constant\\
  $\rho_\bullet$&&internal pebble density                               && $M_p$&&planetesimal mass\\
  $\varSigma_g$&\eq{eq:sigma-model}&gas column density                  && $R_H$&\eq{eq:rhill}&Hill radius\\
  $\varSigma_d$&$Z\varSigma_g$&pebble column density                    && $t_B$&\eq{eq:tbondi}&Bondi time\\
  $k$&\eq{eq:faunsed-plaw}&power law of unsedimented distribution       && $R_B$&\eq{eq:rbondi}&Bondi radius\\
  $p$&\eq{eq:definep}&power law of column density distribution          && $M_t$&\eq{eq:Mt}&transition mass\\
  $q$&\eq{eq:defineq}&power law of internal density                     && $M_{\rm HB}$&\eq{eq:Mthill}&Hill-Bondi transition mass\\
  $D$&\eq{eq:D-coefficient}&Coefficient of column density distribution  && $R$&$\sqrt[3]{\frac{3M_p}{4\pi\rho_\bullet}}$&planetesimal radius\\
  $Z$&$\varSigma_d/\varSigma_g$&Dust-to-gas ratio                       && $v_{\rm esc}$&$\sqrt{\frac{2GM_p}{R}}$&escape velocity\\
  $\rho_\bullet^{(0)}$&&internal density of largest grain               && $\St_p$     &\eq{eq:stp}&Stokes number past planetesimal\\
  $r$&&radial coordinate                                                && $M_{\rm BL}$&\eq{eq:Mtgeo}&Bondi-geometric transition mass\\
  $r_c$&\eq{eq:sigma-model}&cutoff radius                               && $t_{\rm acc}$&\eq{eq:tacc}&accretion time\\
  $W$&\eq{eq:Wada}&column density distribution                          && $h$&$H_g/r$&aspect ratio\\
  $R_{\rm acc}$&\eq{eq:racc}&drag-modified accretion radius             && $m_p$&&characteristic streaming instability mass\\
  $\xi$&\eq{eq:racc}&coefficient                                        && $\psi$&\eq{eq:psi}&shorthand\\
  $\dot{M}$&&mass accretion rate                                        && $T$&\eq{eq:temperature-model}&gas temperature\\
  $c_s$ &$\sqrt{T c_p (\Gamma-1)}$ & sound speed                        && $\Gamma$& & adiabatic index \\
  $\mu$ & & mean molecular weight                                       && $c_p$ & $\frac{R_{\rm gas}}{\mu} \frac{\Gamma}{\left(\Gamma-1\right)}$ & specific heat at constant pressure \\
  $R_{\rm gas}$ & & gas constant                                        && $c_v$ & $c_p/\Gamma$ & specific heat at constant volume\\\hline
\end{tabular}
\end{center}
\end{table*}

\begin{figure*}
  \begin{center}
    \resizebox{\textwidth}{!}{\includegraphics{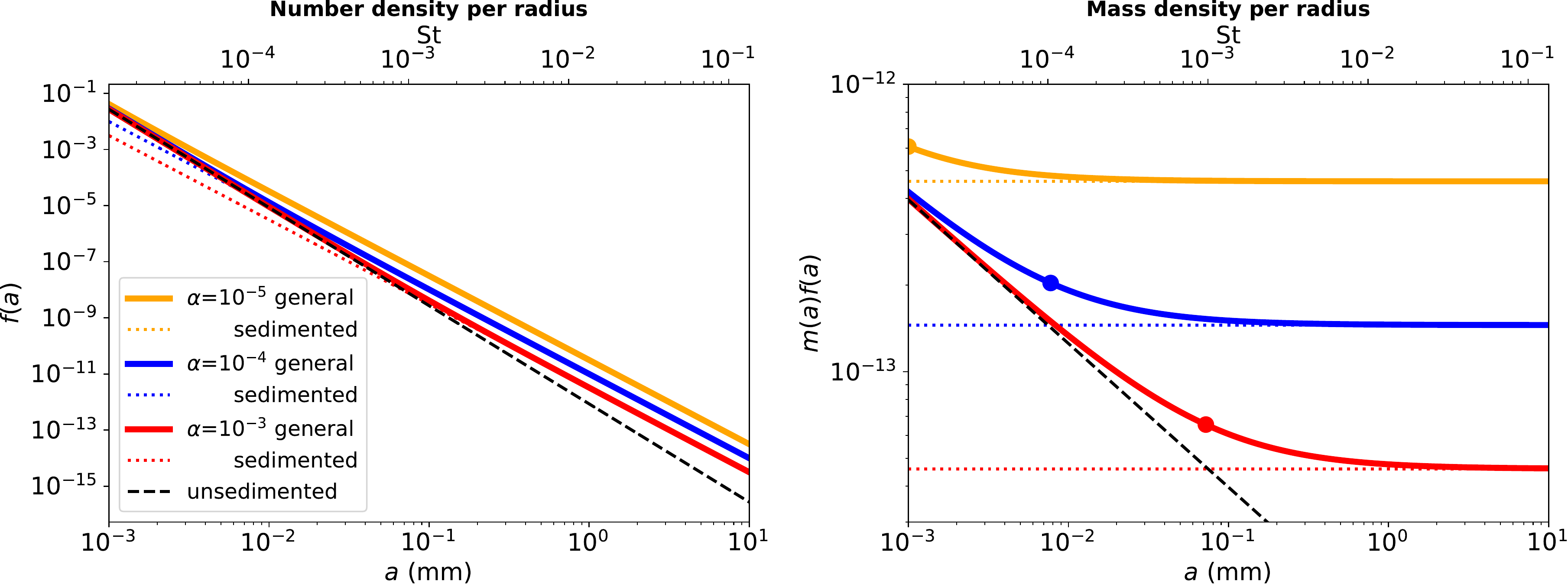}}
\end{center}
\caption{{\it Left:} The grain distribution function $f(a)$ in the
  midplane. Integrated over $a$, this function yields the number density $n$ in
  the midplane. This model is calculated at 20\,AU with
    density and temperature according to \eqs{eq:sigma-model}{eq:temperature-model}, $Z=0.01$,
  constant $\rho_\bullet$, and MRN
  (unsedimented, $\St \ll \alpha$, black dashed line). Three values of $\alpha$ are shown
  (solid lines). The ``sedimented'' limits ($\St \gg \alpha$, dotted
  lines) are shown for comparison. {\it Right:} mass density
  distribution, i.e., the left panel multiplied by the mass of a
  pebble. Integrated, this function yields the grain density $\rho_{d0}$
  in the midplane. The distributions follow the unsedimented line for $\St \lesssim \alpha$,
and the sedimented line for $\St \gtrsim \alpha$, as expected. The
mass function is constant with $a$ in the sedimented limit because of
the MRN choice: the $a^{-3.5}$ power law is canceled by the combination
of the mass of the particle $a^3$ and the extra $\sqrt{a}$ dependency
from the sedimentation. Large dots mark the point where $\St=\alpha$.}
\label{fig:distrofunctions}
\end{figure*}

\section{Distribution Functions}
\label{sect:model}

Consider the grain size distribution

\beq
F(a,z) \equiv \frac{\partial n}{\partial a}
\label{eq:BigF}
\eeq

\noindent that defines the number density $n$; here, $a$ is the grain
radius and $z$ the vertical coordinate. We integrate it to yield

\beq
n(a,z) = \int_{0}^{a} F(a\pa,z) \ da\pa,
\label{eq:numberdensity}
\eeq

\noindent and $n(z)\equiv n(a_{\rm max},z)$. The volume density is found by multiplying $F(a,z)$ by the
mass $m(a)$ of a single grain 

\beq
\rho_d(a,z) = \int_{0}^{a} m(a\pa) \ F(a\pa,z) \ da\pa.
\label{eq:rhoz}
\eeq

\noindent and again, $\rho_d(z)\equiv \rho_d(a_{\rm max},z)$. Due to sedimentation, we can write, for an equilibrium between
  diffusion and gravity \citep{Dubrulle+95}

\beq
F(a,z) \equiv f(a) \ e^{-z^2/2H_d^2}, 
\label{eq:bigF}
\eeq

\noindent defining the function $f(a)$, which is the size distribution function in the
midplane. In \eq{eq:bigF}, $H_d$ is the grain scale height, a function of $a$ \citep{KlahrHenning97,LyraLin13}

\beq
H_d = H_g \ \sqrt{\frac{\alpha}{\St+\alpha}},
\label{eq:Hd}
\eeq

\noindent where $H_g$ is the gas scale height, $\alpha$ is a
dimensionless vertical diffusion parameter{\footnote{This
    parameter is equivalent to the
Shakura-Sunyaev parameter \citep{ShakuraSunyaev73} for isotropic
turbulence of equal diffusion of mass and momentum
\citep{YoudinLithwick07,Yang+18}.}}, and $\St$ is the Stokes number,
a non-dimensionalization of the grain radius, normalized
  by the grain internal density $\rho_\bullet$ and the gas column
  density $\varSigma_g$

\beq
\St \equiv \frac{\pi}{2} \frac{a \ \rho_\bullet}{\varSigma_g}.
\label{eq:stokes}
\eeq

\subsection{The distribution function in the midplane}

To find $f(a)$, consider spherical grains  

\beq
m(a) = \frac{4\pi}{3}a^3 \rho_\bullet
\label{eq:pebblemass}
\eeq

\noindent and the column density 

\beq
\varSigma_d(a) \equiv \int_{-\infty}^{\infty} \rho_d(a,z) \ dz.  
\eeq

\noindent Substituting \eq{eq:rhoz}, and integrating in $z$, we find 

\beq
\varSigma_d (a) = \frac{2^{5/2}\pi^{3/2}}{3}\int_{0}^{a}
\rho_\bullet \ H_d \ a\pa\,^3 \ f(a\pa) \  da\pa,   
\label{eq:varSigma1}
\eeq

\noindent and the total column density $\varSigma_d \equiv \varSigma_d(a_{\rm max})$. We keep the internal density $\rho_\bullet$ inside the integral because
it is in general a function of radius, if grains have different
composition. Given 

\beq
\varSigma_d (a) = \int_0^{a} \frac{\partial \varSigma_d (a\pa)}{\partial a\pa} da\pa,
\label{eq:varSigma2}
\eeq

\noindent we find, equating the integrands of \eq{eq:varSigma1} and
\eq{eq:varSigma2}, and solving for $f(a)$

\beq
f(a) =  \frac{3}{2^{5/2}\pi^{3/2} H_g \rho_\bullet}   \sqrt{1 + \frac{\St}{\alpha}} \ a^{-3} \frac{\partial \varSigma_d (a)}{\partial a}, 
\label{eq:fada0}
\eeq

\noindent where we also substituted \eq{eq:Hd} for $H_d$ as a function of
  $\St$. The distribution is determined if we find an expression for
  $\partial_a\varSigma_d (a)$.

\subsubsection{Sedimented and unsedimented limits}

To find the general solution, we need to find the expression for
$\partial_a \varSigma_d$  in \eq{eq:fada0}. We do so by realizing that
even though the midplane volume density is modified by
sedimentation, the column density $\varSigma_d$ is not. The two limits
of $f(a)$ are, first, the ``sedimented'' limit,  for $\St\gg\alpha$

\beq
f(a)^{({\rm sed})} =  \frac{3}{8 \pi H_g \rho_\bullet^{1/2} \varSigma_g^{1/2}\alpha^{1/2}} \ a^{-2.5} \frac{\partial \varSigma_d (a)}{\partial a},
\eeq

\noindent and, second, the unsedimented limit, for $\St \ll \alpha$

\beq
f(a)^{({\rm unsed})} =  \frac{3}{2^{5/2}\pi^{3/2} H_g \rho_\bullet} \ a^{-3} \frac{\partial \varSigma_d (a)}{\partial a},
\label{eq:faunsed}
\eeq

\noindent where we have substituted the Stokes number given by
\eq{eq:stokes}. Since the column density does not change with
sedimentation, we can find $\partial_a \varSigma_d$ by either
limit.

We assume a power-law dependency for the unsedimented distribution in the midplane

\beq
f(a)^{({\rm unsed})} \ \propto \ a^{-k} 
\label{eq:faunsed-plaw}
\eeq

\noindent where $k$ is a constant (the MRN distribution corresponds to
  $k=3.5$). Equating  \eq{eq:faunsed} and \eq{eq:faunsed-plaw}

\beq
\frac{\partial \varSigma_d (a)}{\partial a} \propto \rho_\bullet \ a^3 \  a^{-k}.
\eeq 

We thus write

\beqn
\frac{\partial \varSigma_d (a)}{\partial a} &\propto& a^{-p};\label{eq:definep}\\
\rho_\bullet &\propto& a^{-q};\label{eq:defineq}\\
p-q &=& k-3.\label{eq:pqk}
\eeqn

\noindent We can then write the column density distribution as a power law

\beq
\frac{\partial \varSigma_d (a)}{\partial a}  = D \ a^{-p}.
\eeq 

\noindent Integrating it in $a$, equating to \eq{eq:varSigma2}, and solving
for the constant $D$, we find

\beq
D = \frac{(1-p)Z\varSigma_g}{a_{\rm max}^{1-p}};
\label{eq:D-coefficient}
\eeq

\noindent here we also substitute $\varSigma_d = Z \varSigma_g$, where $Z$
is the metallicity. Considering now the variation of the internal density 

\beq
\rho_\bullet(a) = \rho_{\bullet}^{(0)} \left(\frac{a}{a_{\rm max}}\right)^{-q},
\eeq

\noindent the full distribution is found at last 

\beq
f(a) =  \frac{3 (1-p)Z\varSigma_g }{2^{5/2}\pi^{3/2} H_g
  \rho_{\bullet}^{(0)} a_{\rm max}^{4-k}}  \sqrt{1 +
  a \frac{\pi}{2} \frac{\rho_\bullet(a)}{\varSigma_g\alpha}} \ a^{-k}.
\label{eq:fada}
\eeq

Notice that to keep $f(a)$ positive definite, the
  solution requires $p<1$. For $q=0$, \eq{eq:pqk}  constrains $k<4$.

  The gas density used is

  \beq
\varSigma_g = 10^{3} \, {\rm g\,cm}^{-2} \ \left(\frac{r}{\rm AU}\right)^{-1}\ e^{-r/r_c}
\label{eq:sigma-model}
  \eeq

\noindent i.e. the self-similar solution to the viscous evolution
equations \citep{Lynden-BellPringle74}. Here $r$ is the distance
  to the star, and $r_c$ a truncation radius. We choose $r_c=$100\,AU. For the temperature, we use the irradiated, radially
  optically thick, vertically optically thin model of \citet[][see also \citealt{Ida+16}]{Kusaka+70}

  \beq
  T = 150 \, {\rm K} \ \left(\frac{r}{\rm AU}\right)^{-3/7} 
  \label{eq:temperature-model}
  \eeq
  
In addition, we assume metallicity $Z=0.01$, adiabatic index
  1.4, and mean molecular weight 2.3.

We plot the resulting distributions in
\fig{fig:distrofunctions}, for 20\,AU, maximum grain size $a_{\rm max} = 1$\,cm, and
$k=-3.5$. For internal density we use $\rho_\bullet^{(0)} =
3.5$\,g\,cm$^{-3}$ and $q=0$. The left panel shows $f(a)$; the right
panel $m(a)f(a)$. The functions are shown for three values of $\alpha$
(solid lines). The unsedimented ($\St \ll \alpha$, black dashed
line) and ``sedimented'' ($\St \gg \alpha$,  dotted lines) limits are shown for comparison. We see that the
sedimented distributions follow the unsedimented line for $\St \lesssim \alpha$,
and the sedimented line for $\St \gtrsim \alpha$, as expected. The flat profile for the sedimented
  cases is due to the MRN exponent, coupled with $\sqrt{\St}$ from the
  sedimentation.

\subsection{Column density}

For completeness, we define the vertically-integrated grain size distribution 

\beq
W(a) \equiv \int_{-\infty}^\infty F(a,z) \ dz  = \sqrt{2\pi} \ H_d \ f(a),
\label{eq:Wa}
\eeq

\noindent so that the pebble column density is 

\beq
\varSigma_d(a) = \int_{0}^{a} m(a\pa) \ W(a\pa) \ da\pa.
\label{eq:columndensity}
\eeq

Substituting \eq{eq:fada} in \eq{eq:Wa}, we find the column density distribution function

\beq
W(a) =  \frac{3(1-p)Z\varSigma_g}{4 \pi  \rho_\bullet^{(0)} a_{\rm max}^{4-k}}   \ a^{-k} ,  
\label{eq:Wada}
\eeq

\noindent which indeed yields $\varSigma_d = Z \varSigma_g$ when
integrated according to \eq{eq:columndensity}.

\section{Pebble Accretion}
\label{sect:pebbleaccretion}

Having found the size distribution function for the pebble density, we
are in position to apply it to pebble accretion. Pebble accretion
is usually split into three regimes of accretion (loosely coupled, 
Bondi, and Hill accretion), each with 2D and 3D limits. We start by
deriving the exact solution for the 2D-3D transition.

\subsection{Exact solution for the monodisperse 2D-3D transition}

The 3D and 2D limits of pebble accretion correspond to whether or not the accretion is embedded, i.e, if the accretion
radius $R_{\rm acc}$ exceeds the height of the pebble column. The quantity governing the
transition is $R_{\rm acc}/H_d$, or rather 

\beq
\xi \equiv \left(\frac{R_{\rm acc}}{2H_d} \right)^2,
\label{eq:xi}
\eeq

\noindent which we will show a posteriori. The monodisperse mass accretion rates in
these limits are \citep{LambrechtsJohansen12}

\beqn
\dot{M}_{\rm 3D} &=& \lim_{\xi \rightarrow 0} \dot{M} =   \pi R_{\rm acc}^2 \rho_{d0} \delta v, \label{eq:monodisperse-3d}\\
\dot{M}_{\rm 2D} &=& \lim_{\xi \rightarrow \infty} \dot{M} = 2R_{\rm acc} \varSigma_d \delta v, \label{eq:monodisperse-2d}
\eeqn

\noindent where $\delta v$ is the velocity at which the pebble approaches the
accretor, and $\rho_{d0}$ is the midplane density. In principle, we could apply \eq{eq:rhoz} with \eq{eq:fada} on \eq{eq:monodisperse-3d}; and \eq{eq:columndensity}
with \eq{eq:Wada} on \eq{eq:monodisperse-2d}, working with the two limits
separately. Yet, given that $\xi$ is a function of grain size, and
there are other transitions to deal with (loose coupling/Bondi/Hill), it is preferable to work with a general
expression for $\dot{M}$, which we derive in this section. 

Considering parallel horizontal chords of infinitesimal thickness in the vertical direction until
the full accretion radius is taken into account, the general expression for the mass accretion rate is 

\beq
\dot{M} = \int_{-R_{\rm acc}}^{R_{\rm acc}} 2 \sqrt{R_{\rm acc}^2-z^2} \ \rho_{d0} \  {\rm  exp}\left(-\frac{z^2}{2H_d^2}\right) \delta v \ dz.
\eeq

\noindent Following \cite{Johansen+15} we define the stratification integral

\beq
S \equiv \frac{1}{\pi R_{\rm acc}^2} \int_{-R_{\rm acc}}^{R_{\rm acc}}
2 \sqrt{R_{\rm acc}^2-z^2} \ {\rm
  exp}\left(-\frac{z^2}{2H_d^2}\right) \ dz, 
\eeq

\noindent so that the mass accretion rate is generalized into one expression as

\beq
\dot{M} = \pi R_{\rm acc}^2 \rho_{d0} \ S \ \delta v. 
\eeq

While \cite{Johansen+15} use a square approximation for the accretion
radius, we find the exact solution of the stratification integral

\beq
S = e^{-\xi}\left[ I_0(\xi) +  I_1(\xi) \right],
\label{eq:stratification-analytical}
\eeq

\noindent where $I_\nu(\xi)$ are the  modified Bessel functions of the
first kind, and $\xi$ is given by \eq{eq:xi}. The exact monodisperse accretion rate is 

\beq
\dot{M} = \pi R_{\rm acc}^2 \rho_{d0} \, \delta v \, e^{-\xi}\left[ I_0(\xi) +  I_1(\xi) \right].
\label{eq:monodisperse-general}
\eeq

\begin{figure}
  \begin{center}
    \resizebox{\columnwidth}{!}{\includegraphics{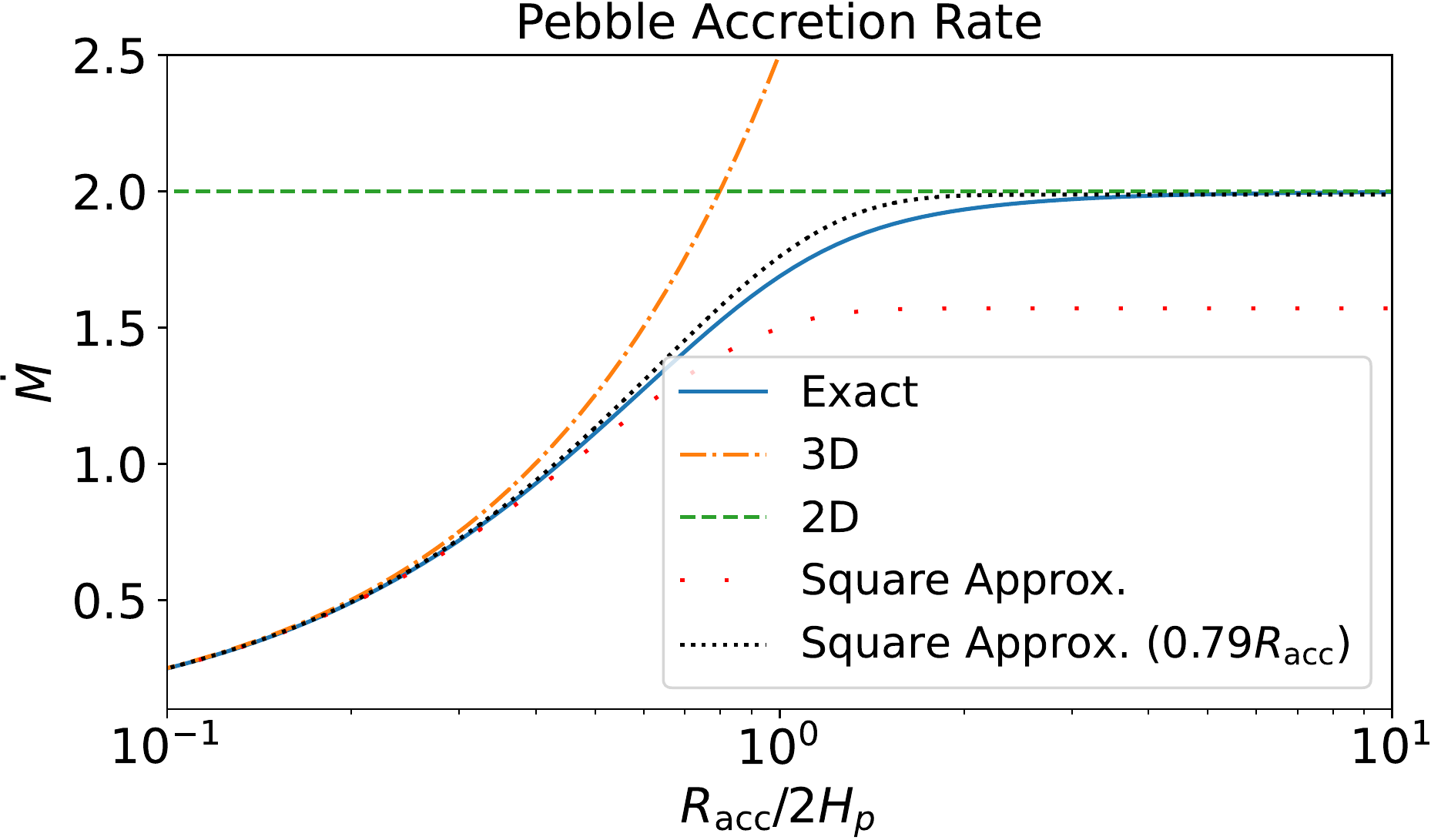}}
\end{center}
\caption{General expression for the monodisperse pebble accretion rate
  \eqp{eq:monodisperse-general}. The 3D and 2D limits
  (eqs~\ref{eq:monodisperse-3d}
  and~\ref{eq:monodisperse-2d}, respectively) are recovered. The square approximation is
    also shown for comparison.}
\label{fig:monodisperse-general}
\end{figure}

Indeed for $\xi\rightarrow 0$, the Bessel functions tend to $ I_0(0) =
1$, $I_1(0)=0$, and we recover 3D accretion \eqp{eq:monodisperse-3d}.  For  $\xi\rightarrow
\infty$, both Bessel functions tend to $e^{\xi}/\sqrt{2\pi \xi}$, and 2D
accretion is recovered \eqp{eq:monodisperse-2d}. \fig{fig:monodisperse-general} shows the
agreement graphically. The square approximation is shown for comparison.

\subsection{Polydisperse prescription}

To generalize \eq{eq:monodisperse-general} into a polydisperse
description, we consider the integrated polydisperse accretion rate to
be $\dot{M} \equiv \dot{M}
(a_{\rm max})$, where

\beq
\dot{M}(a) = \int_{0}^{a} \frac{\partial \dot{M}(a^\prime)}{\partial a\pa}da\pa, 
\label{eq:integrated-acc-rate}
\eeq

\noindent with

\beq
\frac{\partial \dot{M}(a)}{\partial a }  = \pi R_{\rm acc}^2(a) \ \delta v(a) \ S(a) \ m(a) \ f(a).
\label{eq:mass-accretion-rate-integrand}
\eeq

Indeed, \eq{eq:integrated-acc-rate} with the integrand given by
\eq{eq:mass-accretion-rate-integrand} is equivalent to
\eq{eq:monodisperse-general} if 

\beq
\int_0^{a_{\rm max}} R_{\rm acc}^2(a) \ \delta v(a) \ S(a) \ m(a) \
f(a)  da = \bar{R}_{\rm acc}^2 \ \overline{\delta v} \ \bar{S}
  \, \rho_{d0},
\label{eq:equivalence}
\eeq

\noindent where the overline denotes that the quantity is an ``effective'' quantity,
  independent of pebble size. If the accretion radius $R_{\rm acc}$, the approach velocity
  $\delta v$,  and the stratification integral $S$ were independent of
  the grain radius $a$, \eq{eq:equivalence} would
  be exactly equivalent to replacing the midplane dust density by the
integrated grain size distribution

\beq
\rho_{d0} = \int_0^{a_{\rm max}} m(a) f(a) da,
\label{eq:rhod0-midplane}
\eeq
  
\noindent which is intuitive. We can now use much of the formalism of pebble
  accretion already derived in the literature. The approach velocity $\delta v$ is given by 

\beq
\delta v \equiv \Delta v + \varOmega R_{\rm acc},
\label{eq:deltav}
\eeq

\noindent where $\Delta v$ is the sub-Keplerian velocity reduction
and $\varOmega$ is the Keplerian frequency. The accretion radius is \citep{OrmelKlahr10}

\beq
R_{\rm acc} \equiv \hat{R}_{\rm acc} {\rm exp}\left[-\chi (\tau_f/t_p)^\gamma \right],
\label{eq:racc}
\eeq

\noindent where $\tau_f=\St/\varOmega$ is the pebble friction
  time, $\chi=0.4$ and $\gamma=0.65$ are empirically-determined coefficients, and

\beq
t_p \equiv \frac{GM_p}{\left(\Delta v + \varOmega R_H\right)^3}
\label{eq:tp}
\eeq

\noindent is the characteristic passing time scale. Here $G$ is
  the gravitational constant, $M_p$ the mass of the planetesimal, and
  $R_H$ its Hill radius

\beq
R_H \equiv \left(\frac{GM_p}{3\varOmega^2}\right)^{1/3}.
\label{eq:rhill}
\eeq

The variable $\hat{R}_{\rm acc}$ depends on the accretion regime. For
Hill accretion it is 

\beq
\hat{R}_{\rm acc}^{\rm (Hill)}  = \left(\frac{\St}{0.1}\right)^{1/3} R_H,
\label{eq:rhathill}
\eeq

\noindent and for Bondi accretion it is

\beq
\hat{R}_{\rm acc}^{\rm (Bondi)}  = \left(\frac{4\tau_f}{t_B}\right)^{1/2} R_B,
\label{eq:rhatbondi}
\eeq

\noindent where

\beq
R_B \equiv \frac{GM_p}{\Delta v^2}
\label{eq:rbondi}
\eeq

\noindent is the Bondi radius and

\beq
t_B \equiv \frac{R_B}{\Delta v}
\label{eq:tbondi}
\eeq

\noindent is the Bondi time. The transition mass between Bondi
  and Hill accretion is defined by \citep{Ormel17}

\beq
M_{HB} = \frac{M_t}{8\St},
\label{eq:Mthill}
\eeq

\noindent where

\beq
M_t \equiv \frac{\Delta v^3}{G\varOmega}.
\label{eq:Mt}
\eeq

A third regime also exists, of accretion of loosely coupled pebbles, for
which the accretion radius is the physical radius $R$ augmented by the
gravitational focusing cross-section

\beq
R_{\rm acc}^{\rm (geo)} = R \sqrt{1+\frac{v_{\rm esc}^2}{\Delta v^2}},
\label{eq:rgeo}
\eeq

\noindent where $v_{\rm esc}$ is the escape velocity of the planetary seed. In
this regime the grains are so loosely coupled they behave
almost like planetesimals, except for small enough grains, that
remain coupled to the gas and follow the gas streamlines. The quantity that defines this
latter transition is \citep{Ormel17}

\beq
\Stp = \frac{\Delta v \ \tau_f}{R},
\label{eq:stp}
\eeq

\noindent that is, the friction time normalized by the time to pass past the
planetesimal; a planetesimal Stokes number (hence the ``p'' in $\Stp$). For $\Stp < 1$, we set $R_{\rm acc}^{\rm (geo)}=0$. The
transition mass $M_{BL}$ between Bondi and 
loosely coupled accretion happens at \citep{Ormel17}

\beq
M_{BL} = \frac{M_t}{8} \St.
\label{eq:Mtgeo}
\eeq

\begin{figure*}
  \begin{center}
    \resizebox{\columnwidth}{!}{\includegraphics{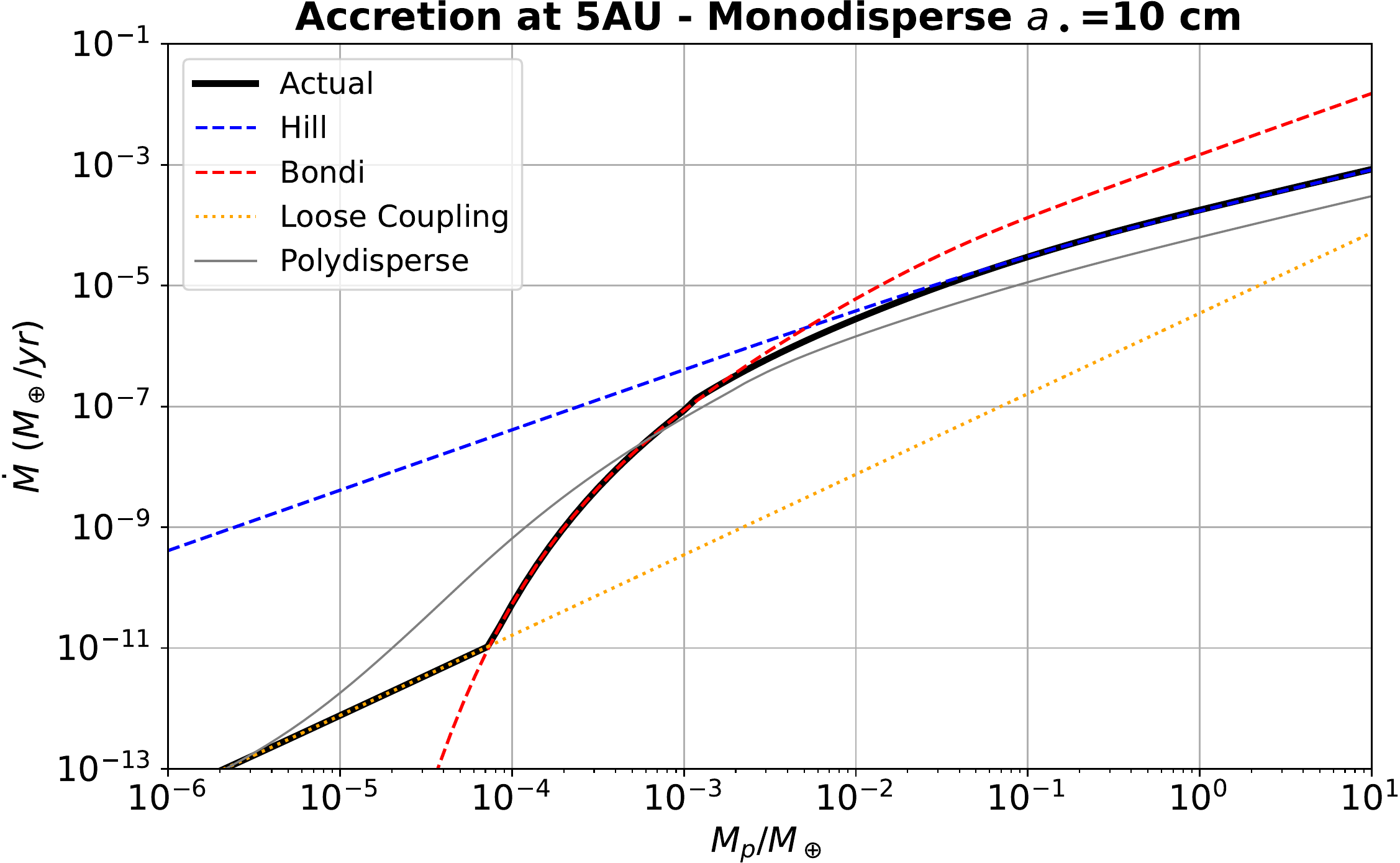}}
    \resizebox{\columnwidth}{!}{\includegraphics{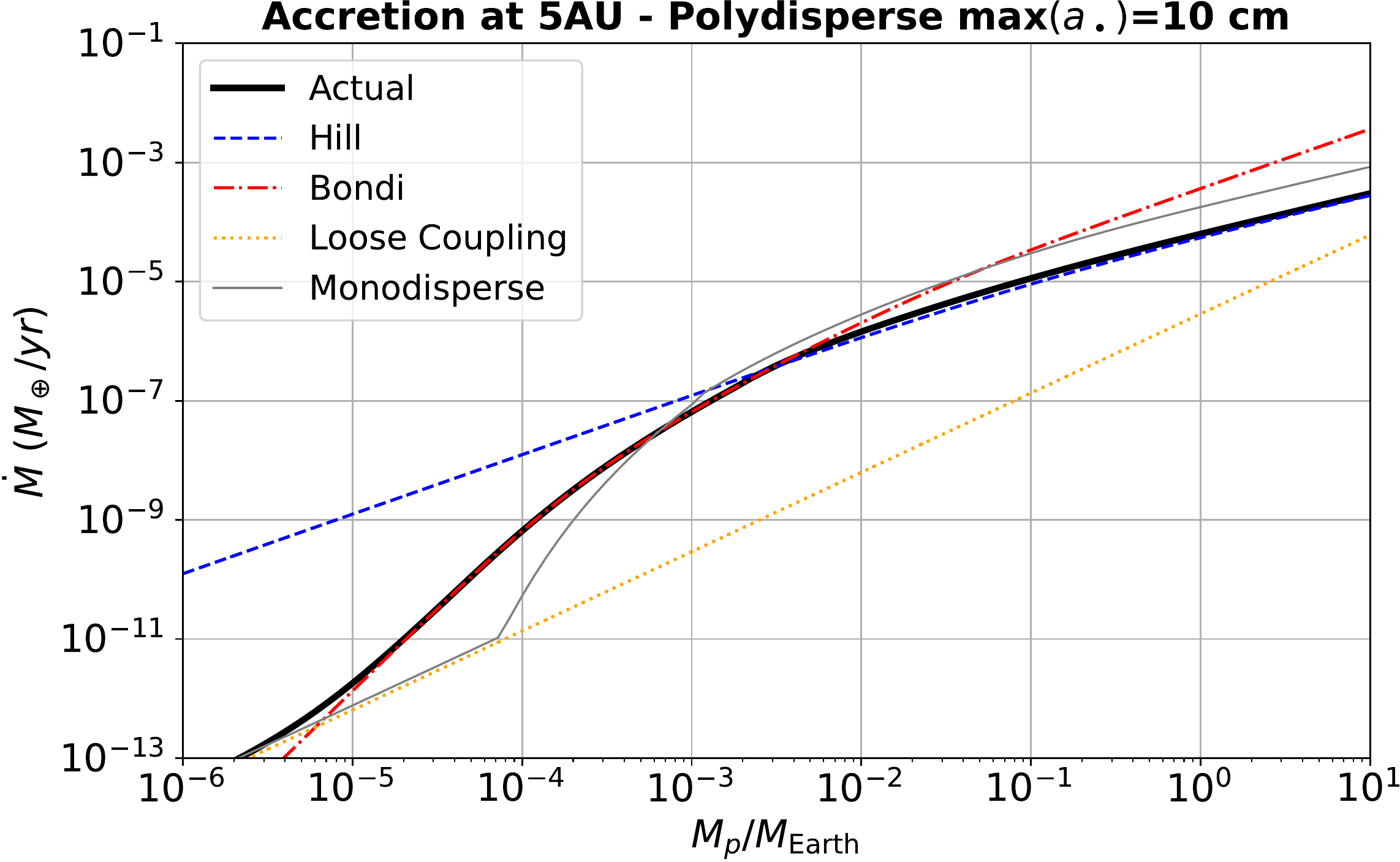}}
\end{center}
\caption{Comparison between monodisperse (left) and the integrated polydisperse
  (right) accretion rates \eqp{eq:integrated-acc-rate}. The left panel uses the parameters of Fig.\,4
  of \citet{JohansenLambrechts17}, except that we use the monodisperse
general equation here derived \eqp{eq:monodisperse-general}. A pebble
size of 10\,cm is not supported by observations but we keep this size
for benchmarking purposes. The polydisperse accretion rate is
  reproduced in the left plot, and the monodisperse accretion rate in the right plot (grey lines), for comparison. The Hill accretion yields a lower
accretion rate (3/7 lower than monodisperse) because other pebbles sizes are present,
not only $a=10$\,cm. The main difference is the accretion rate
for polydisperse Bondi accretion being up to two orders of magnitude more efficient than
  monodisperse, and the onset of pebble accretion happening over one order
  of magnitude lower in mass. This occurs because the best-accreted pebble is not present in the
monodisperse distribution, and $a_{\rm max}$ is too loosely
coupled, accreting poorly. Notice the smooth transition from Bondi to Hill
accretion with the exact 2D-3D transition.}
\label{fig:accretion_sizedistro_integrated}
\end{figure*}

\begin{figure*}
  \begin{center}
    \resizebox{\columnwidth}{!}{\includegraphics{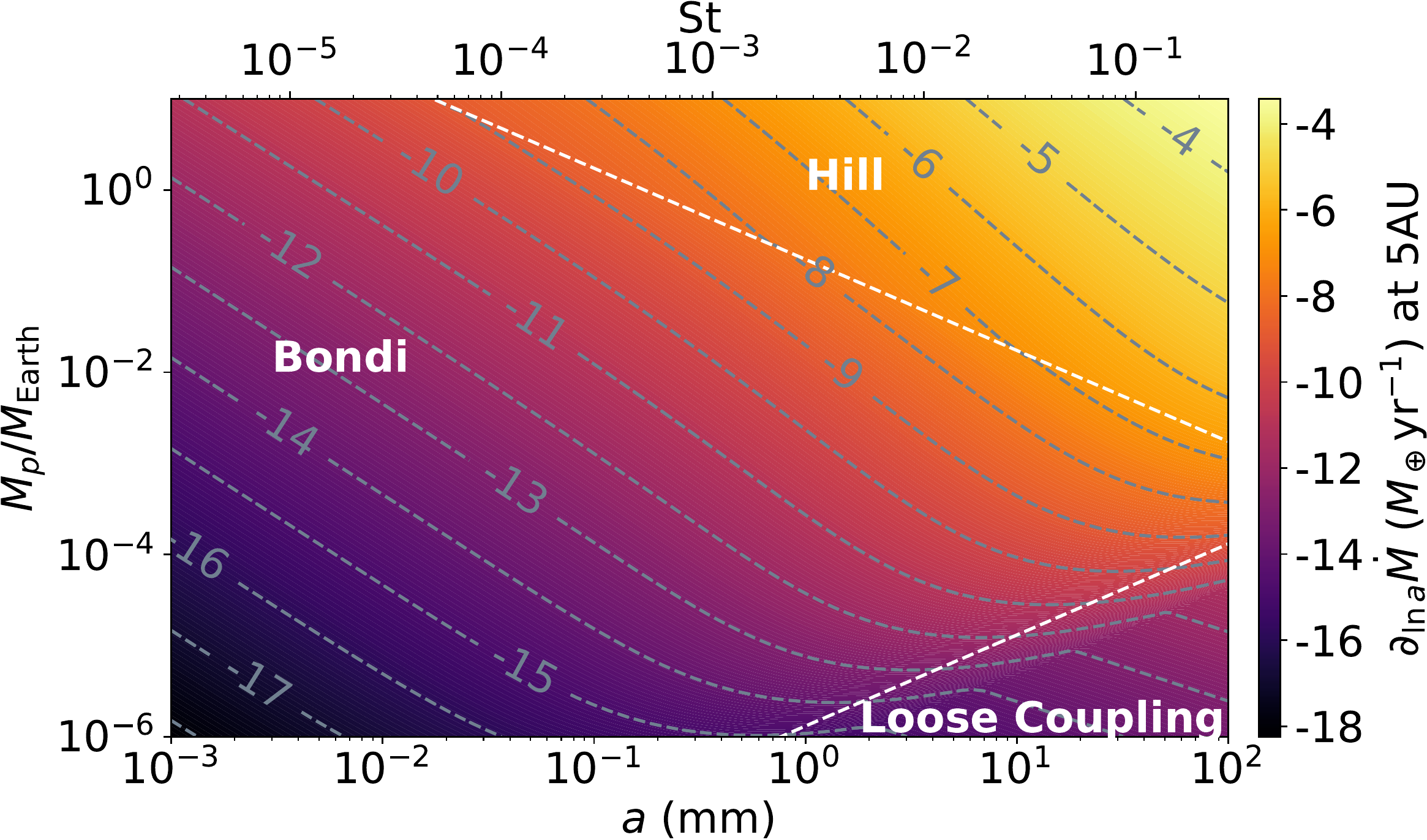}}
    \resizebox{\columnwidth}{!}{\includegraphics{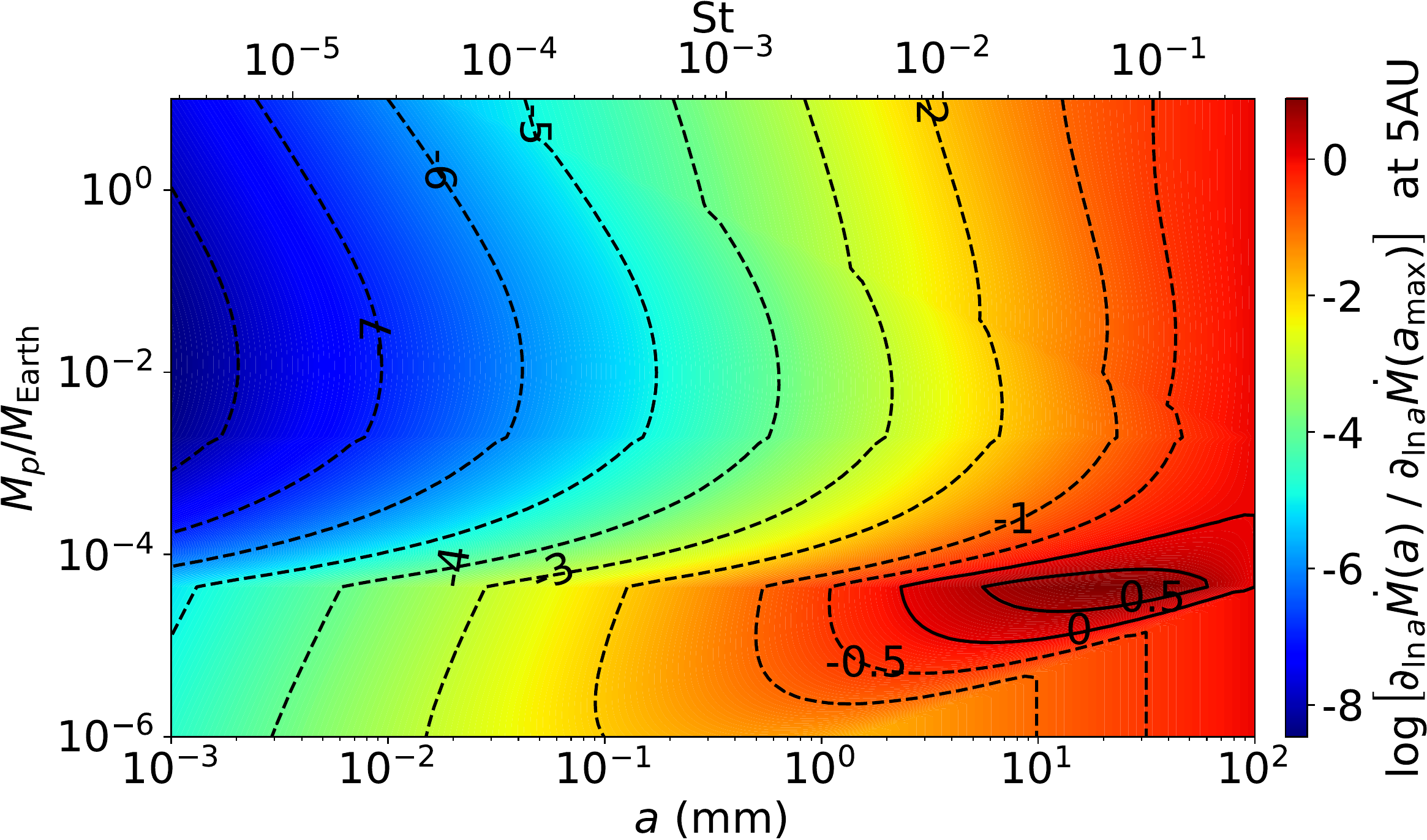}}
  \end{center}
\caption{{\it Left:} The polydisperse pebble accretion rate
  $\partial_{\,\ln\,a}\,\dot{M}$ \eqp{eq:mass-accretion-rate-integrand}, as a
  function of grain radius. In the Hill accretion regime the largest pebble
  present dominates the mass accretion rate. Conversely,
  for Bondi accretion, we see that at a given seed mass the differential accretion
  rate is non-monotonic with grain size. For low enough seed masses, the
  biggest grains, although dominating the mass distribution, accrete
  in the loosely coupled regime. {\it Right:} Same as the left plot, but normalized by the
   accretion rate for $a_{\rm max}$ (proxy for monodispersive). The bright red contours are the
    regions were polydisperse accretion is enhanced over
    monodisperse. We see that it mostly corresponds to the region
    where monodisperse is in the loosely coupled regime, but
    polydisperse is already in Bondi. The best accreted pebbles are
    those for which the stopping time $\tau_f$ equals the Bondi time
    $t_B$. Absent in the monodispersive description, these pebbles may
    contribute less to the mass budget, but
    their enhanced accretion ends up dominating the mass accretion
    rate.}
\label{fig:PebbleAccretion_vs_GrainRadius}
\end{figure*}

\subsection{Polydisperse vs Monodisperse}
\label{sect:results}

We show in the left panel of \fig{fig:accretion_sizedistro_integrated} a reproduction of
the monodisperse accretion rates from \citet{JohansenLambrechts17}, for $a =$ 10\,cm, and at
5\,AU. Even though the observations do not support the existence of
these large grains, we use it for benchmark purposes. The different lines show the pebble accretion rates in the Hill
and Bondi regimes, as well as the
loosely coupled regime for low masses.

The Hill limit (blue dashed line) is recovered for \eq{eq:monodisperse-general} with $\hat{R}_{\rm acc}$ given by
  \eq{eq:rhathill}, and $\delta v =\varOmega R^{(\rm Hill)}_{\rm acc}$. The
  Bondi limit (red dashed line) is recovered for
  \eq{eq:monodisperse-general} with $\hat{R}_{\rm acc}$ given by
  \eq{eq:rhatbondi}, and $\delta v = \Delta v + \varOmega R^{(\rm
    Bondi)}_{\rm  acc}$. The actual solution (black thick line) uses

\beq
\hat{R}_{\rm acc} = \left\{
\begin{array}{ll}
\hat{R}_{\rm acc}^{(\rm Hill)} & {\rm if} \ M \geq M_{\rm HB},\\ 
\hat{R}_{\rm acc}^{(\rm Bondi)} & {\rm if} \ M < M_{\rm HB},
\end{array} 
\right.  
\label{eq:racchatgeneral}
\eeq

\noindent and the general $\delta v$ given by \eq{eq:deltav}. The
  mass accretion rate is then the maximum between this and the loosely
  coupled accretion rates.  The loosely
  coupled regime is given by \eq{eq:monodisperse-general} with
  $\delta v=\Delta v$  and $R_{\rm acc}$ given by \eq{eq:rgeo}
  if $\Stp \geq 1 $,  and zero otherwise.

The right panel of \fig{fig:accretion_sizedistro_integrated} shows how the accretion rates differ
when we include a particle size distribution. In this panel we are
showing the integrated accretion rate $\dot{M} \equiv \dot{M}
(a_{\rm max})$ given by \eq{eq:integrated-acc-rate}. The monodisperse line is shown for comparison.

\subsubsection{Slightly lower efficiency in the Hill regime} 

From comparing the plots in \fig{fig:accretion_sizedistro_integrated},
we see that the polydisperse accretion rate is slightly
lower in the regime of Hill accretion; this
occurs because, in the Hill regime, there is less mass at the biggest
pebble size $a_{\rm max}$
compared to monodisperse (where all pebbles are of 10\,cm). We
  work out in \sect{sect:analytical} this reduction factor to be
  exactly 3/7.

\subsubsection{Significantly higher efficiency in the Bondi regime} 

In the Bondi regime, conversely, there are now pebbles to accrete of
friction time similar to the Bondi time. In the monodisperse
regime there were only the 10\,cm pebbles that, for very low mass seed,
behave like infinite $\St$ and do not accrete well. As a result, in the polydisperse case, Bondi accretion is more efficient than
loosely coupled accretion over a wider range of low seed
masses. At the mass where monodisperse experiences the onset of
  pebble accretion (about $10^{-4} M_\oplus$), the polydisperse
  distribution is well into the Bondi regime, which is about 100$\times$
  more efficient. We also see that the onset of pebble accretion
  occurred between $10^{-6}$ and $10^{-5} M_\oplus$, i.e., between
  100-200\,km. This is a significant early onset of pebble
  accretion, that may eliminate the need for planetesimal accretion to bridge the gap between the
  largest masses formed by streaming instability and the onset of
  efficient pebble accretion \citep{Johansen+15,Schafer+17,Li+19}.

We plot in \fig{fig:PebbleAccretion_vs_GrainRadius} the differential
mass accretion rate as a function of pebble size (horizontal axis) and
seed mass (vertical axis). The left panel shows the polydisperse
  mass accretion rate $\partial_{\,\ln\,a}\,\dot{M}$, and the right panel shows the ratio between
  that and the same quantity for the largest grain size in the
    distribution, which we take as a proxy for monodisperse. The three accretion regimes are
labeled in the left plot; one sees the smooth transition between Hill and Bondi
accretion, and the discontinuous transition from Bondi to
loosely coupled. It is seen that, at a given mass, Hill accretion is
monotonic with particle size, but Bondi accretion is not. A
local maximum of mass accretion rate occurs, corresponding to the size for which $\tau_f
= t_{ \rm Bondi}$, which in turn leads to a linear dependency on the
best accreted particle size for a given seed mass. The bright
  red parts of the right plot show where Bondi accretion is more
  efficient than monodisperse. It is the more efficient
  accretion of these grains that boosts the Bondi accretion rates in
  the polydisperse case. We see that it corresponds chiefly to the
  region of the parameter space for which monodisperse accretion was
  in the loosely coupled regime, but the polydisperse is well within
  Bondi. This confirms that indeed it is the accretion of the
  smaller, Bondi-optimal, pebbles, that is increasing the accretion
  rate.

\subsection{Effect of distance}

\begin{figure*}
  \begin{center}
    \resizebox{.33\textwidth}{!}{\includegraphics{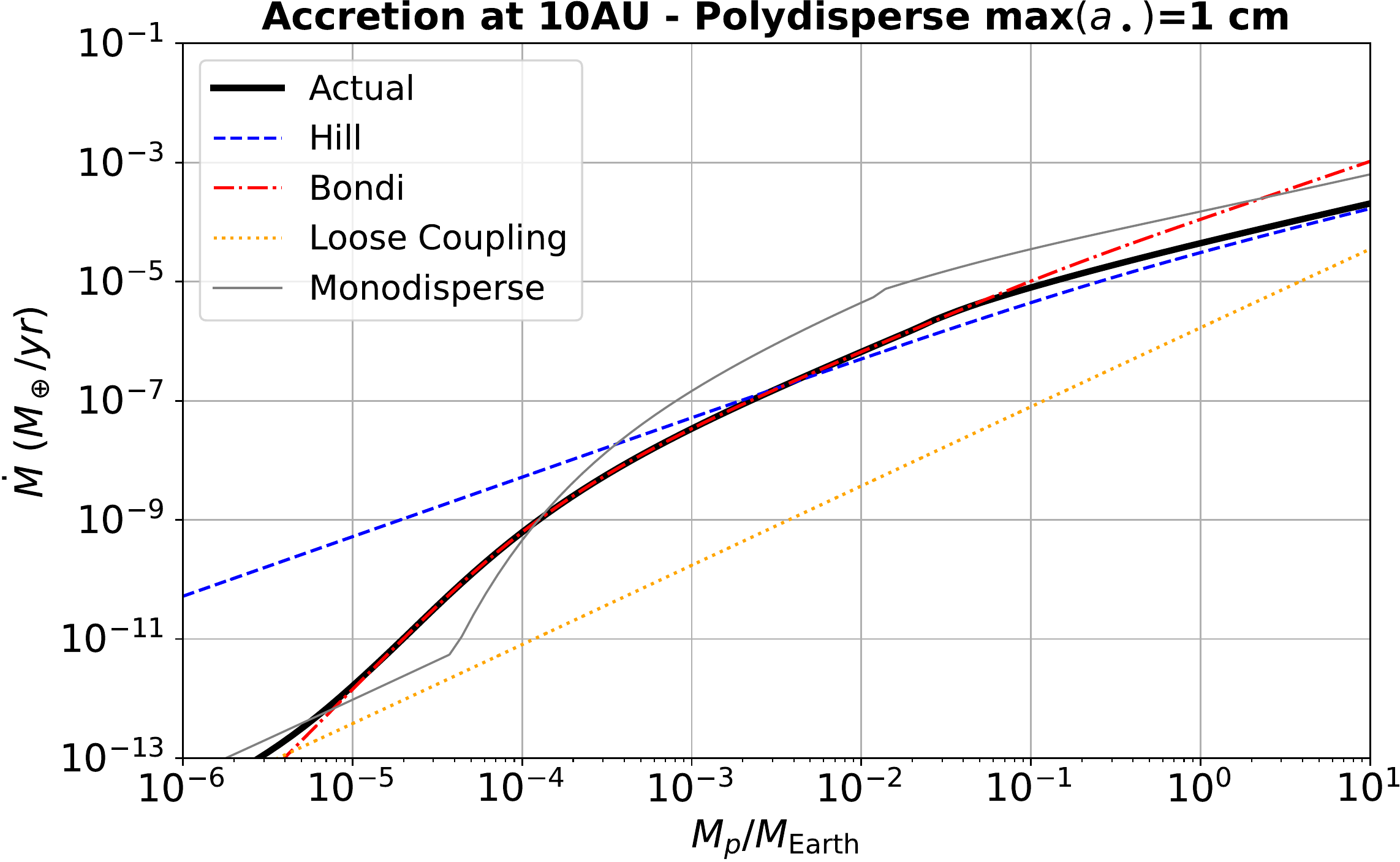}}
    \resizebox{.33\textwidth}{!}{\includegraphics{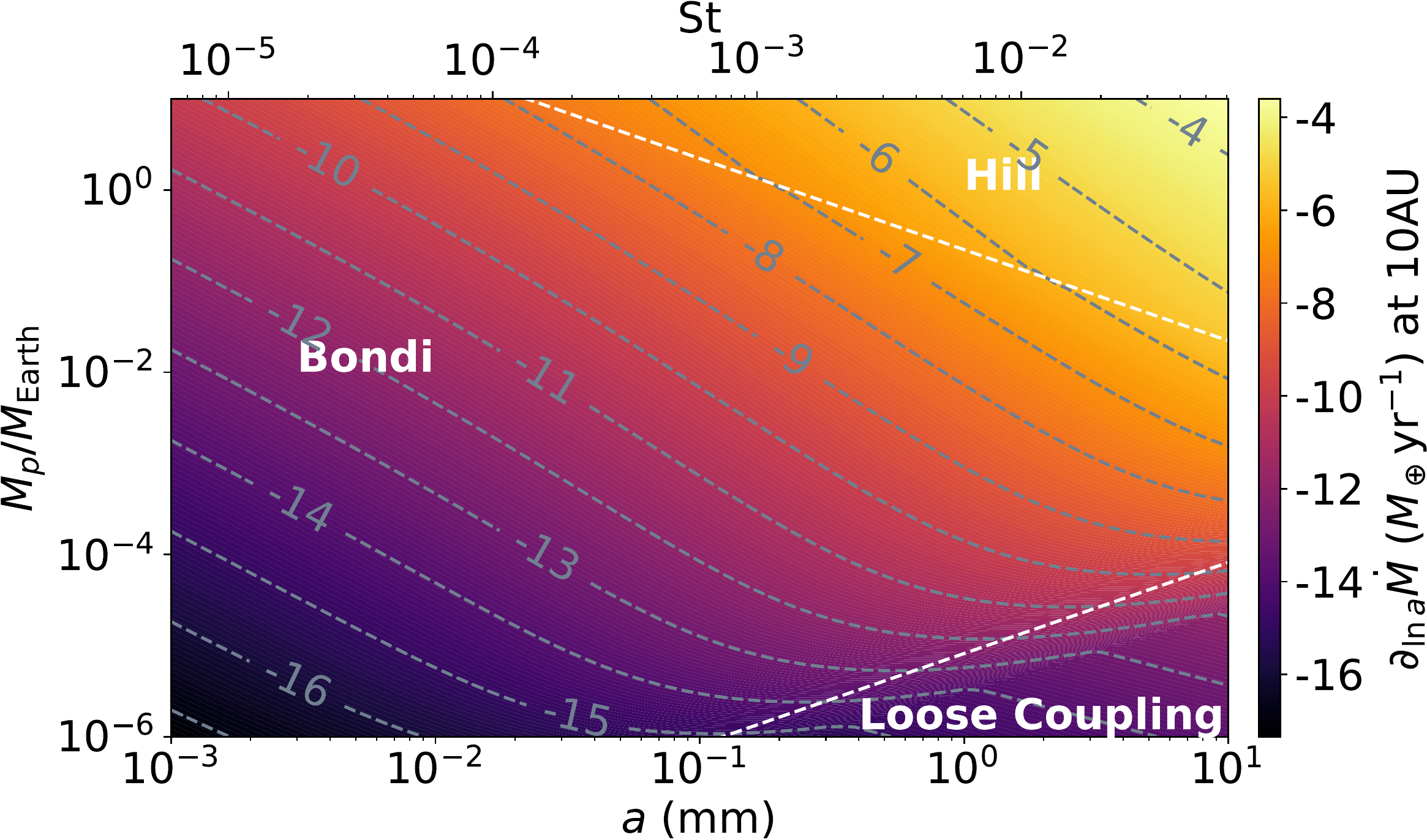}}
    \resizebox{.33\textwidth}{!}{\includegraphics{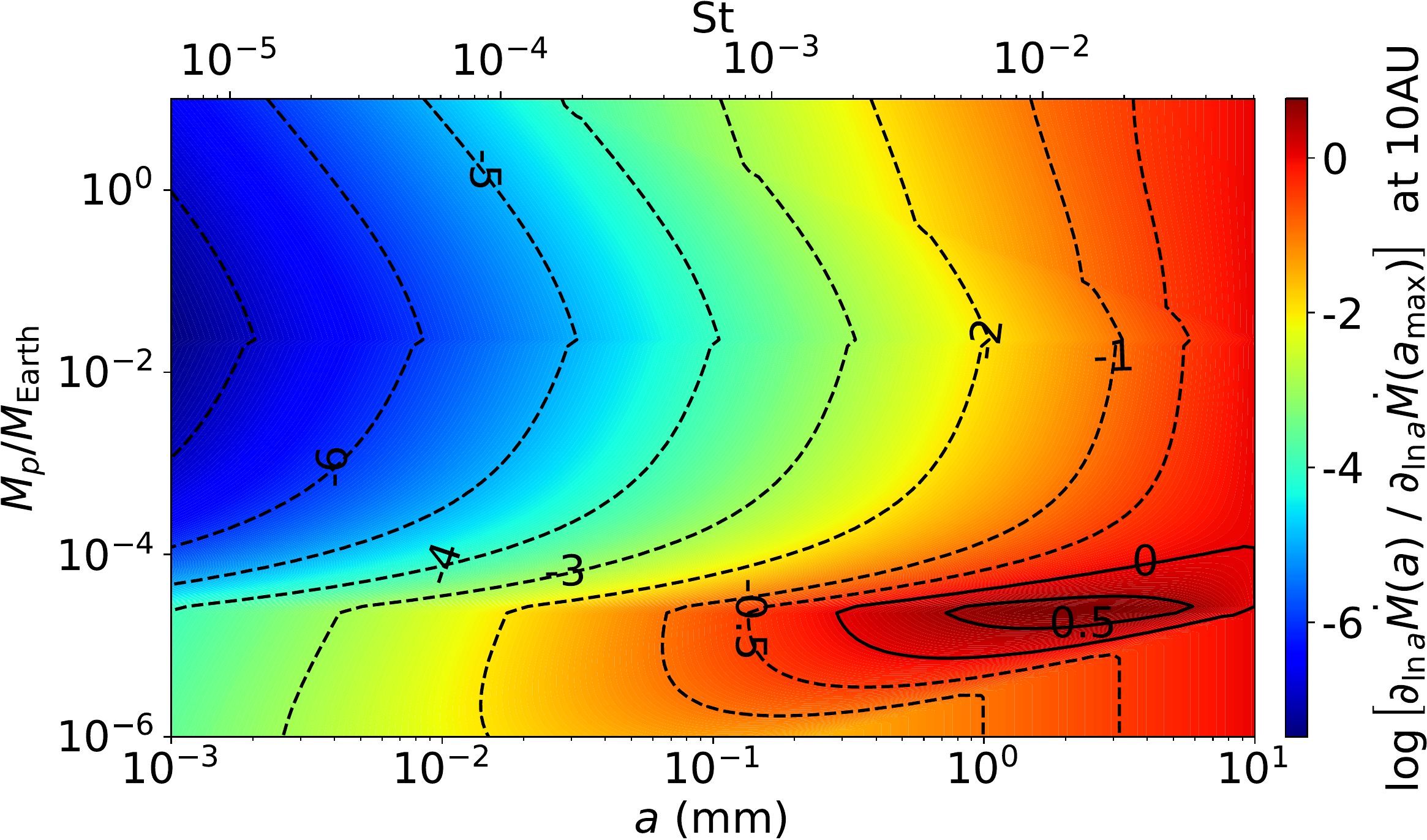}}

    \resizebox{.33\textwidth}{!}{\includegraphics{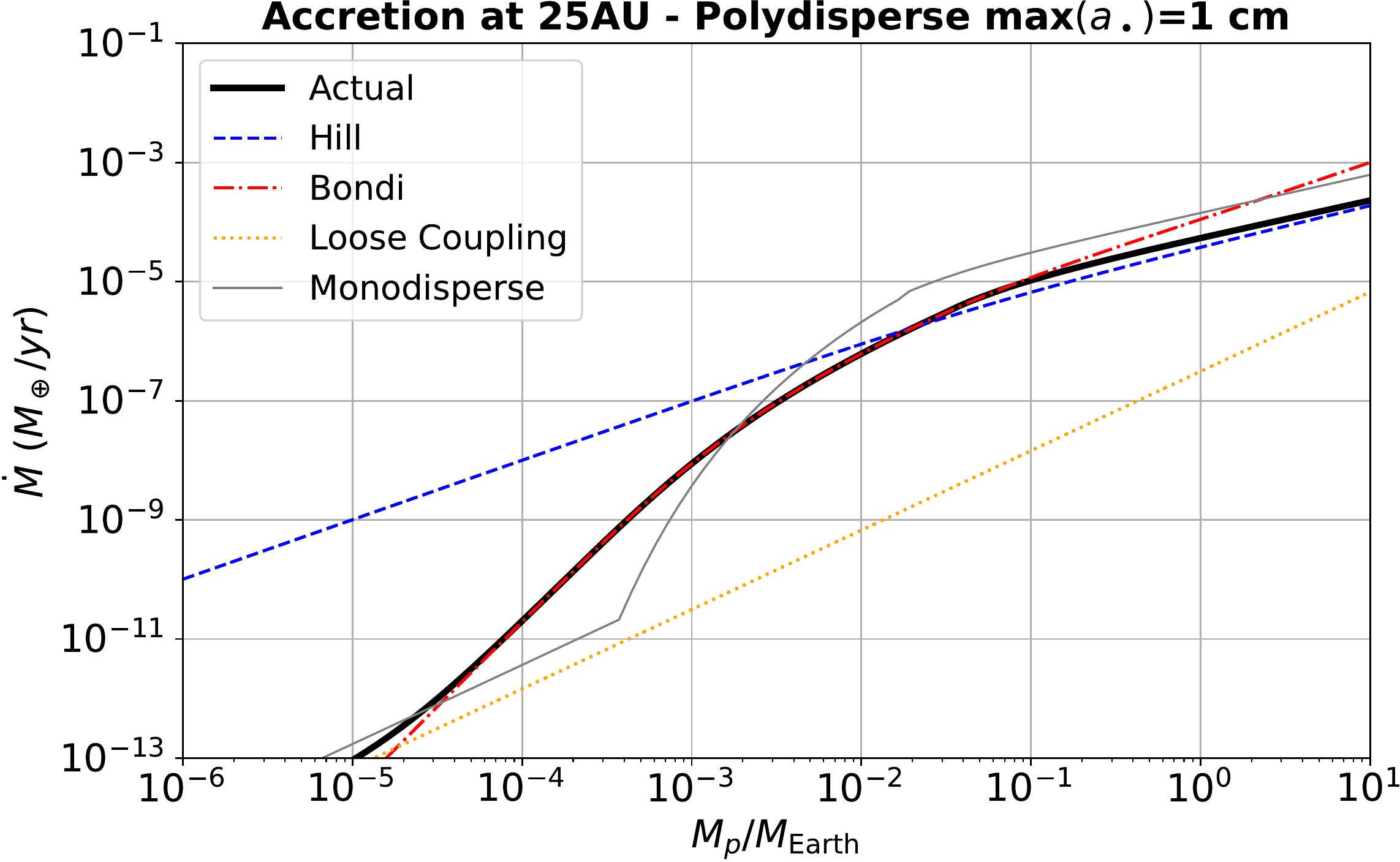}}
    \resizebox{.33\textwidth}{!}{\includegraphics{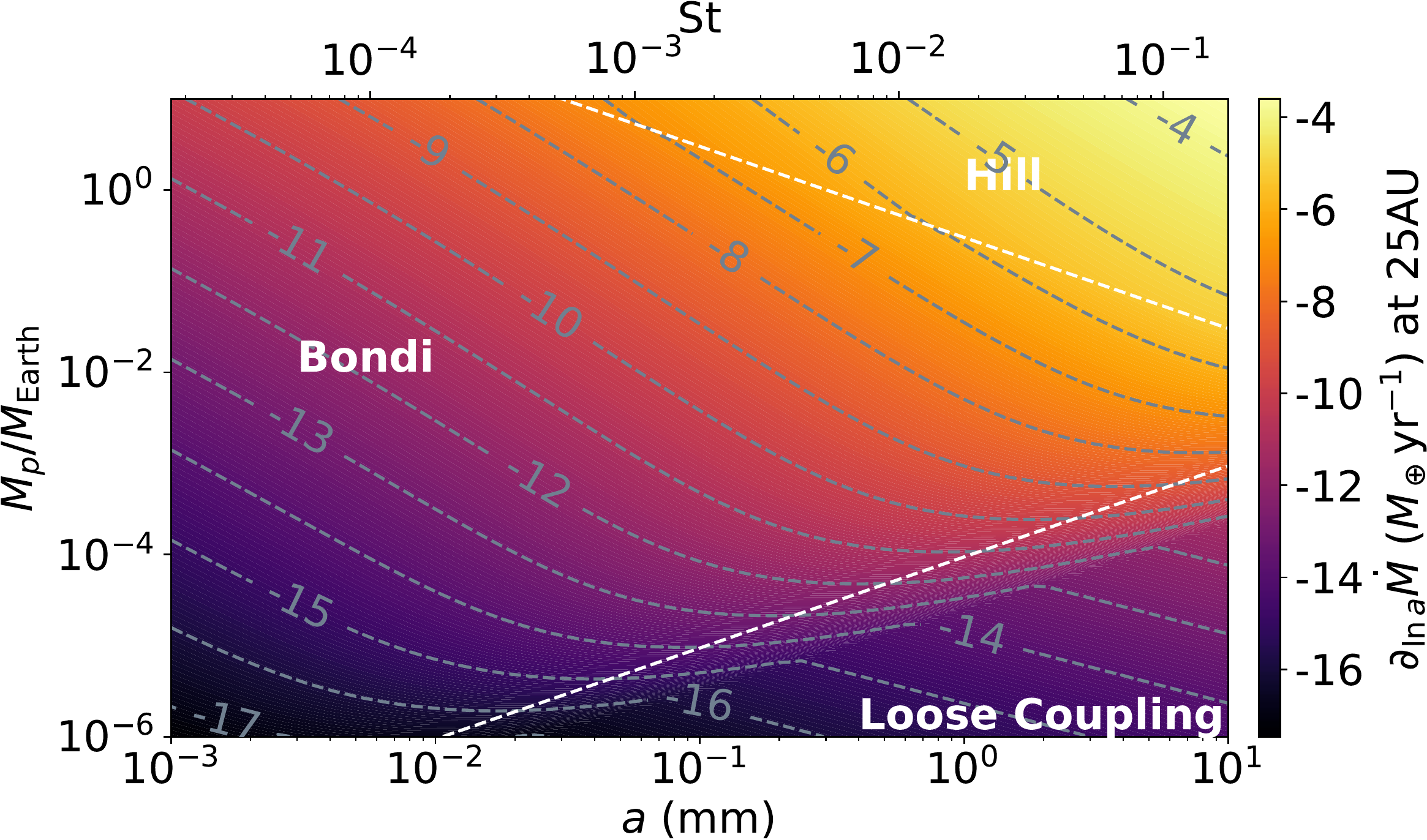}}
    \resizebox{.33\textwidth}{!}{\includegraphics{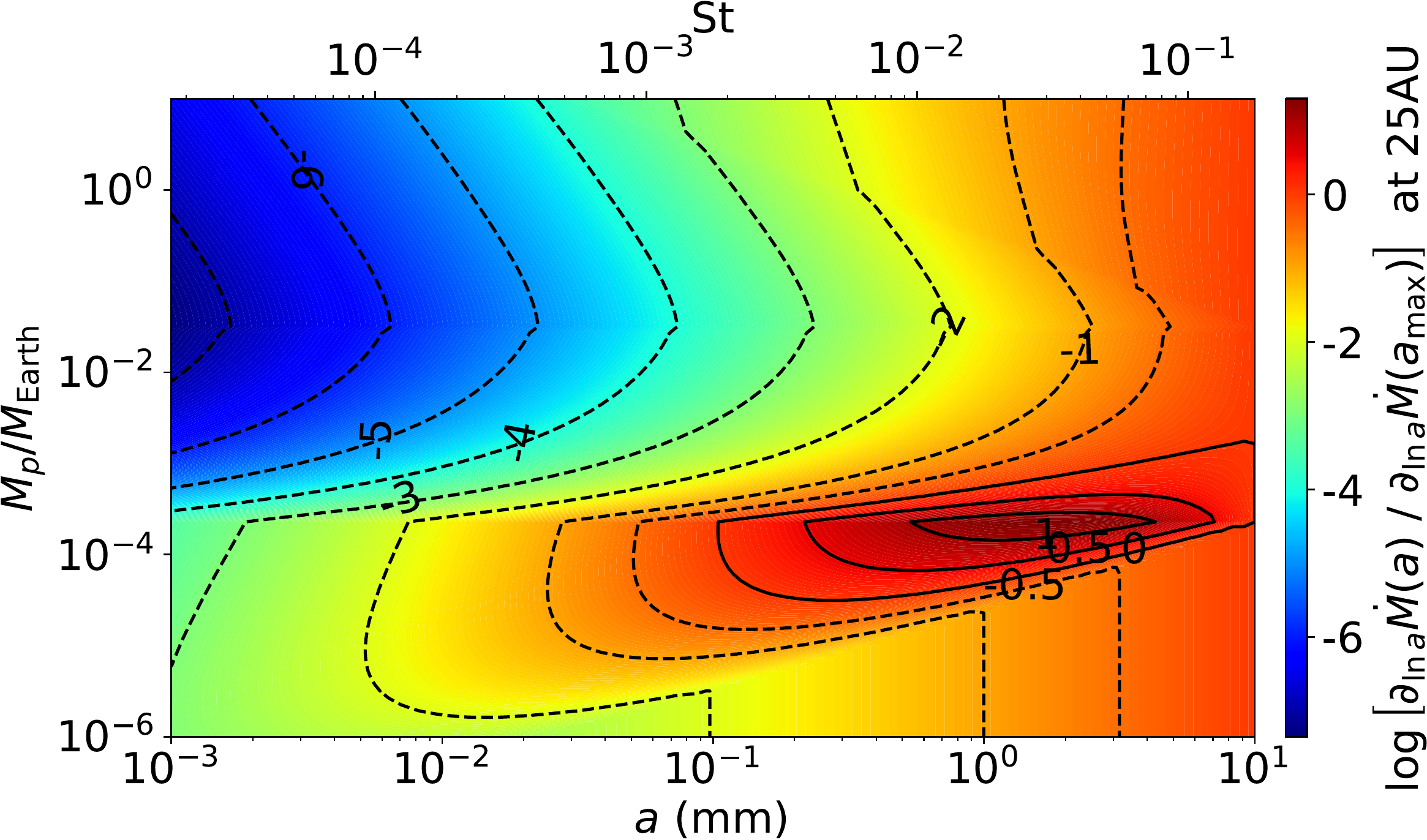}}
    
    \resizebox{.33\textwidth}{!}{\includegraphics{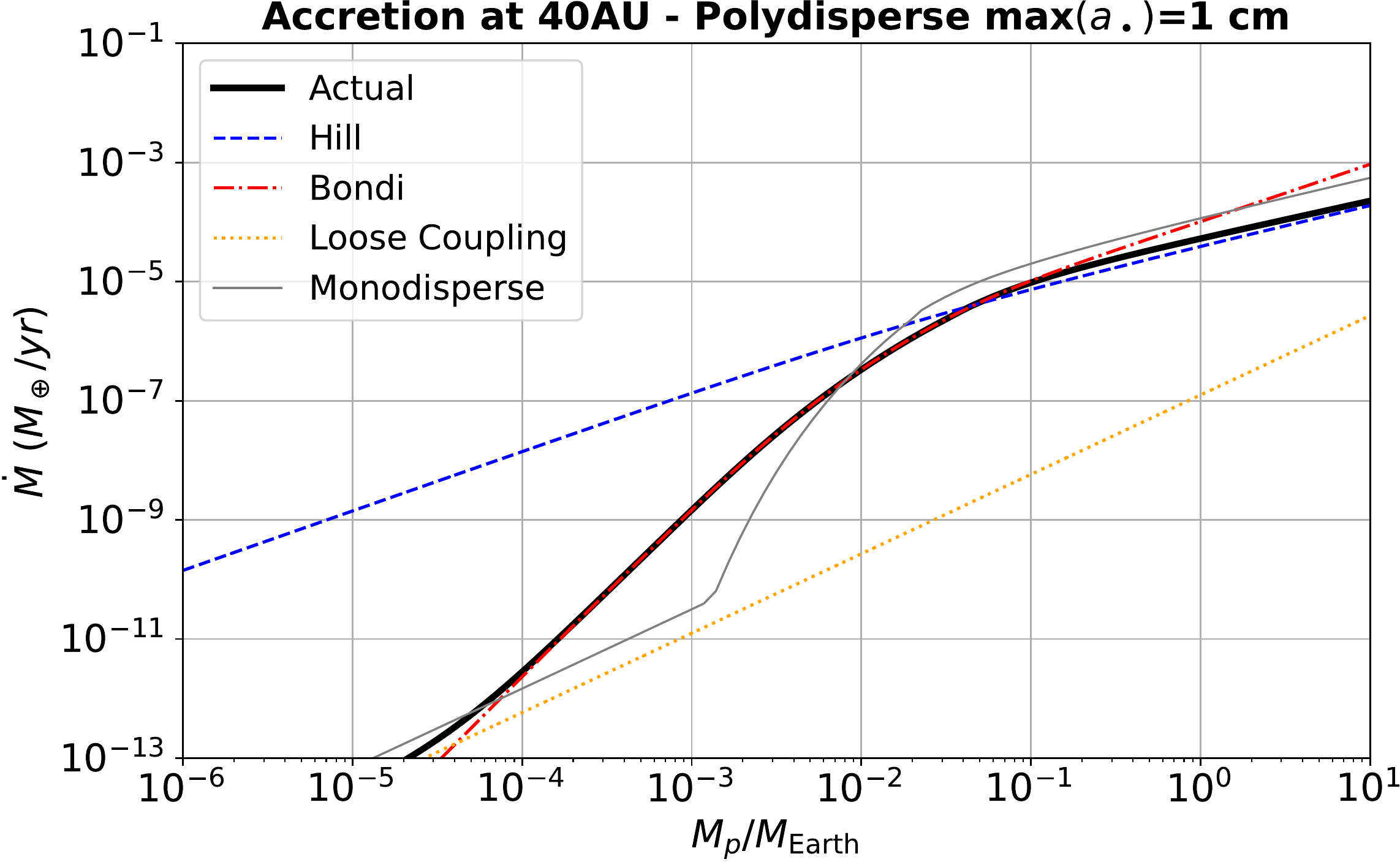}}
    \resizebox{.33\textwidth}{!}{\includegraphics{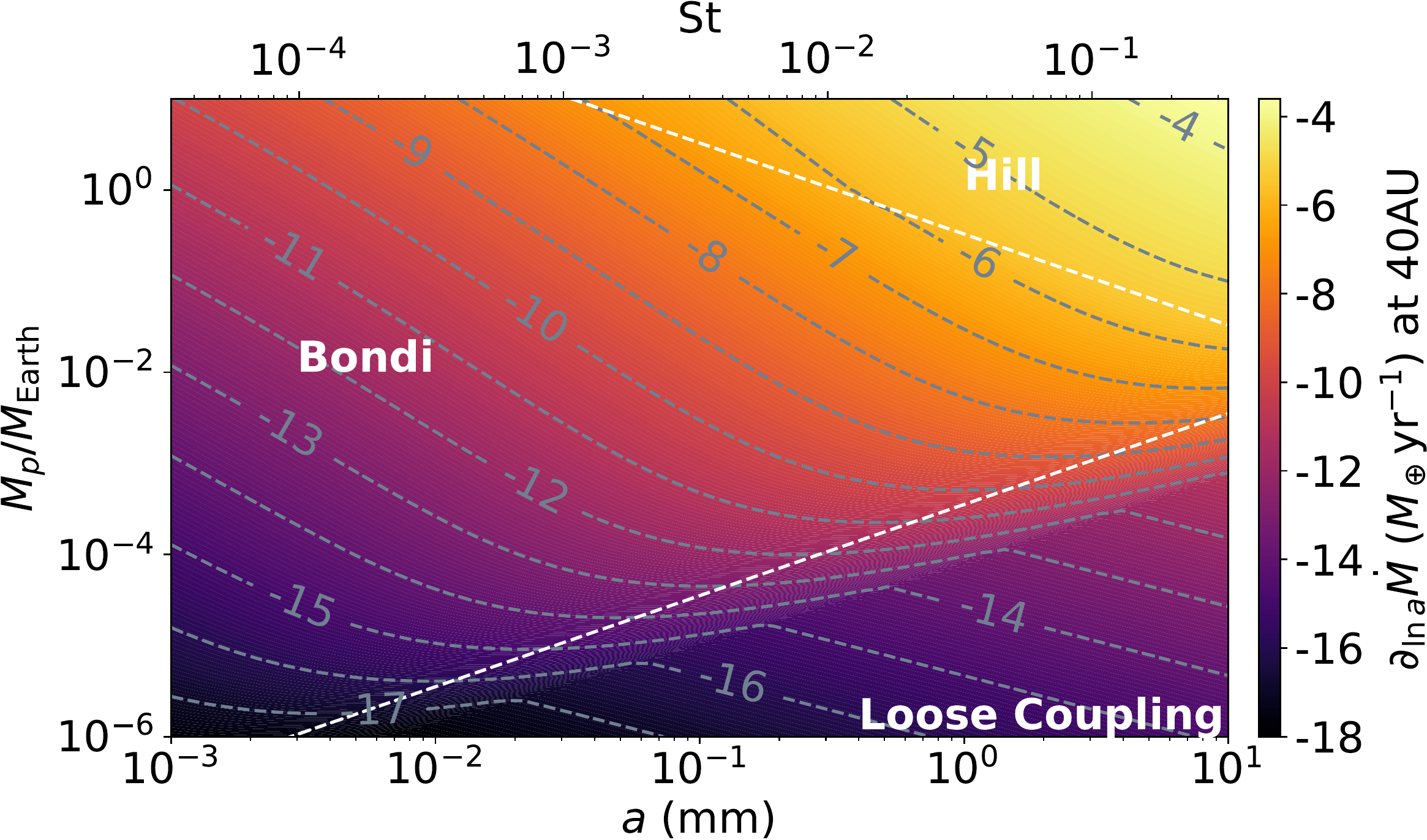}}
    \resizebox{.33\textwidth}{!}{\includegraphics{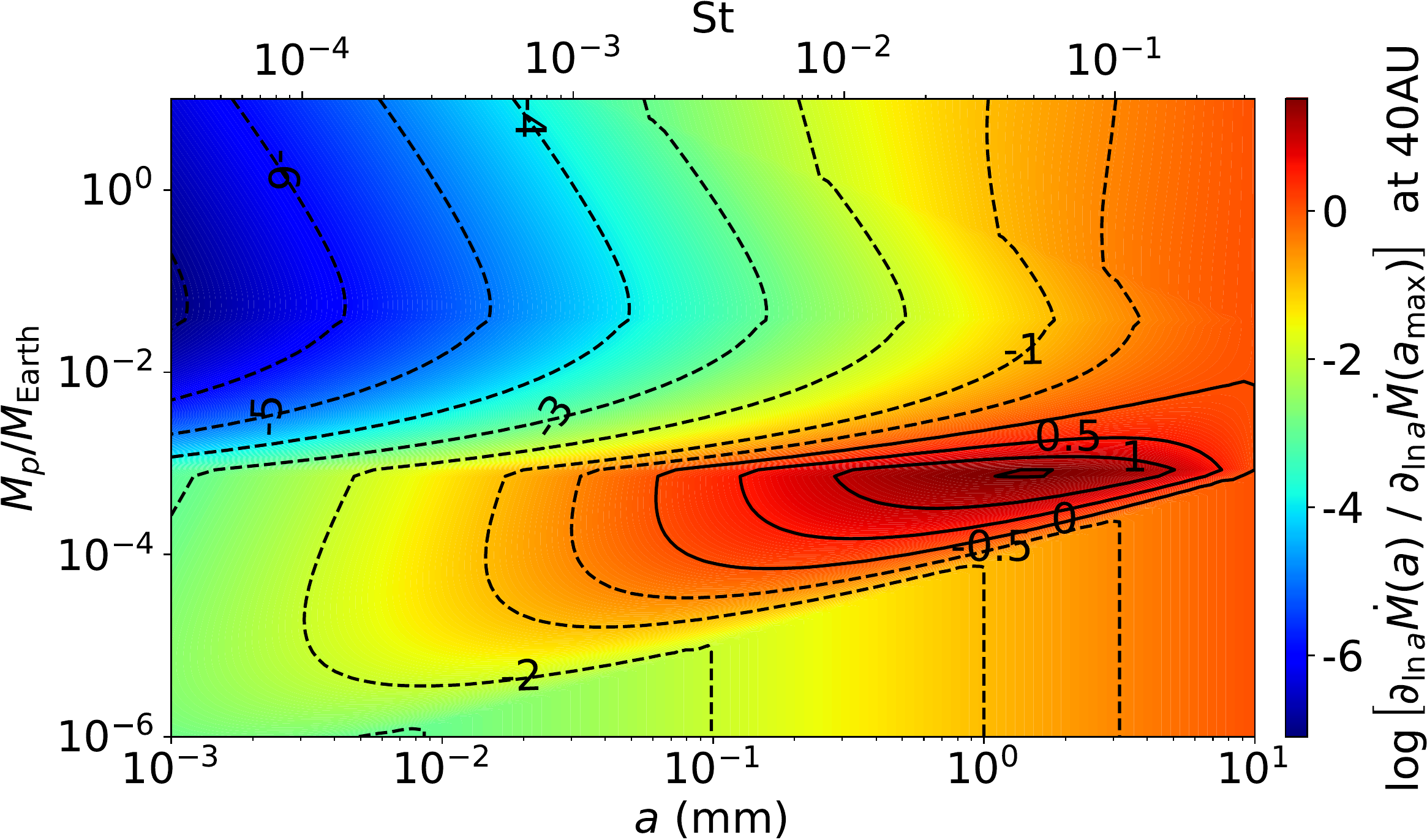}}    
  \end{center}
  \caption{{\it Left:} Same as \fig{fig:accretion_sizedistro_integrated}, right
    plot, but for the density and temperature of
      \eqs{eq:sigma-model}{eq:temperature-model}, Z=0.01, and at different distances. Hill accretion is
    not much affected by distance, but Bondi accretion becomes increasingly less
    efficient as distance increases. Yet, the general trend remains,
    of polydisperse pebble accretion being 1-2
    orders of magnitude more efficient than monodisperse at maximum,
    and showing an earlier onset in mass also by 1-2 orders of magnitude. {\it Middle and
      Right:}  same as \fig{fig:PebbleAccretion_vs_GrainRadius}, at
    difference distances. The pebble size that maximizes Bondi
    accretion decreases as distance increases. This has interesting
    implications, because in the outer disk, the seeds, presumably
    icy, should accrete small grains, presumably silicates. This
    implies the possibility a two-mode formation of Kuiper belt
    objects: icy planetesimal produced by streaming instability of
    larger grains, followed by pebble
    accretion of smaller, silicate, grains.}
\label{fig:PebbleAccretion_vs_GrainRadius_distance}
\end{figure*}

\begin{figure*}
  \begin{center}
    \resizebox{.49\textwidth}{!}{\includegraphics{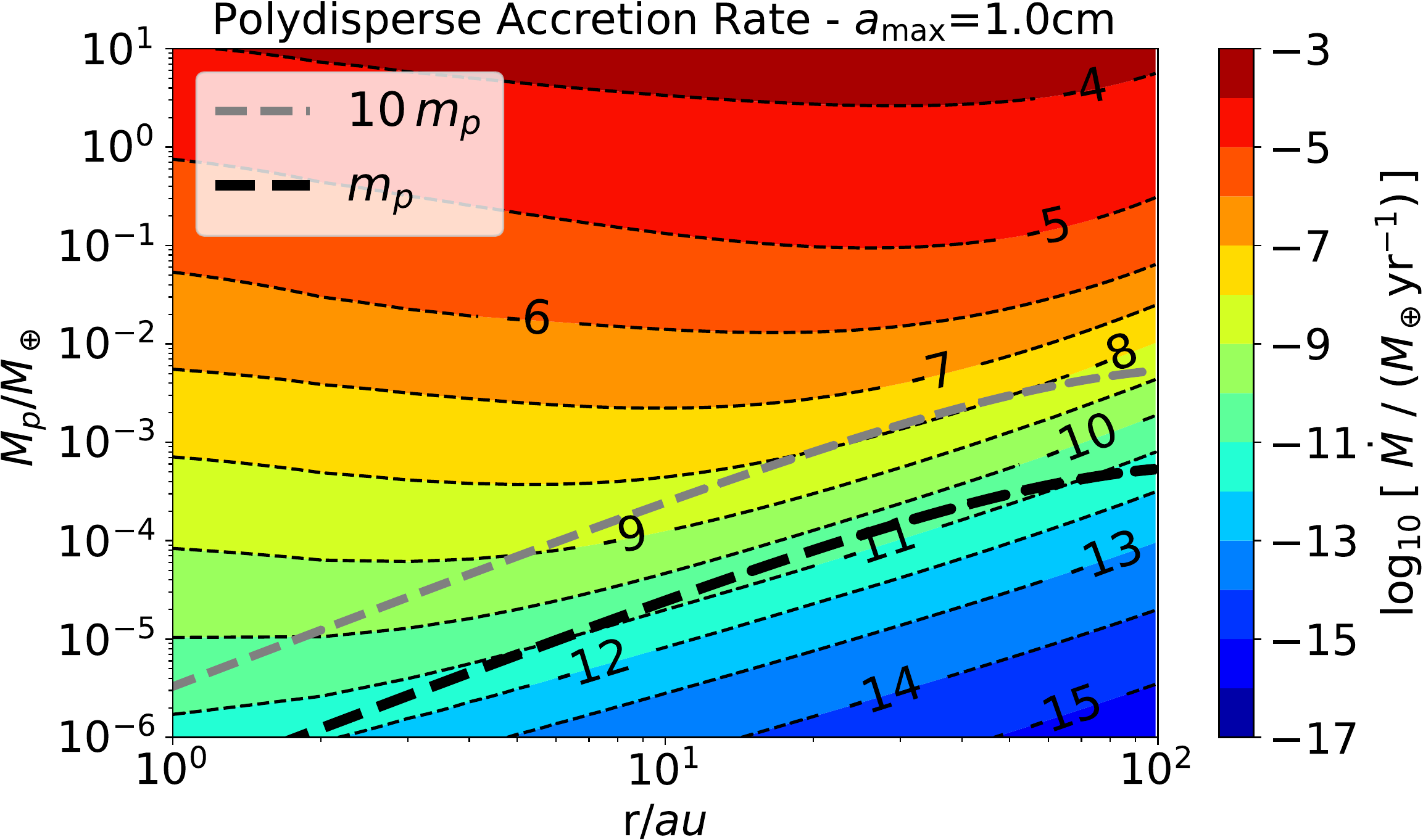}}
    \resizebox{.49\textwidth}{!}{\includegraphics{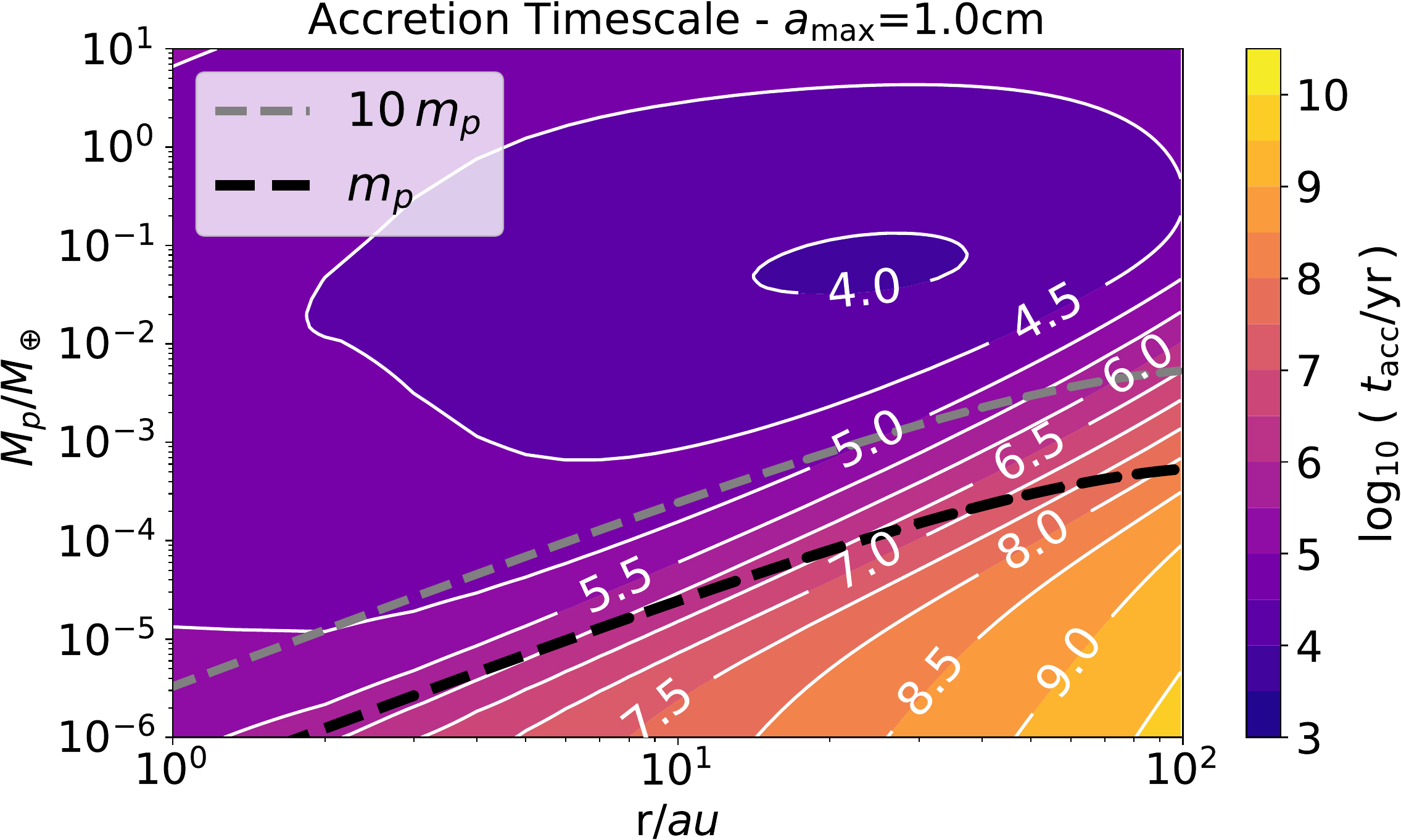}}
  \end{center}
  \caption{{\it Left}: Integrated polydisperse pebble mass accretion
    rate, as a function of distance. The model  uses the density and temperature of
      \eqs{eq:sigma-model}{eq:temperature-model}, Z=0.01,  and
    $\rho_\bullet$ constant. The thick black dashed line shows the
      characteristic size of the planetesimals formed by streaming
      instability \citep{Liu+20,LorekJohansen22}; the grey line
        represents bodies of 10$\times$ the typical mass.  {\it Right:} Accretion times $M/\dot{M}$
      for the same model.  The contour of 6.5 (3Myr) marks the boundary
      where accretion during the lifetime of the nebula is 
      feasible by pebble accretion, without the need for planetesimal accretion. That contour
      corresponds to 3\,AU, 10\,AU, and 30\,AU, for $10^{-6} M_\oplus$,
      $2\times 10^{-5} M_\oplus$, and $2\times 10^{-4}
      M_\oplus$, respectively. These masses correspond to 100\,km
      radius, $10^{-2}$ and $10^{-1}$ Pluto masses,
      respectively. The typical products of streaming
        instability have $<$3 Myr growth times up to 30\,AU.}
\label{fig:IntPebbleAccretion_vs_GrainRadius_distance}
\end{figure*}

\begin{figure*}
  \begin{center}    
    \resizebox{.33\textwidth}{!}{\includegraphics{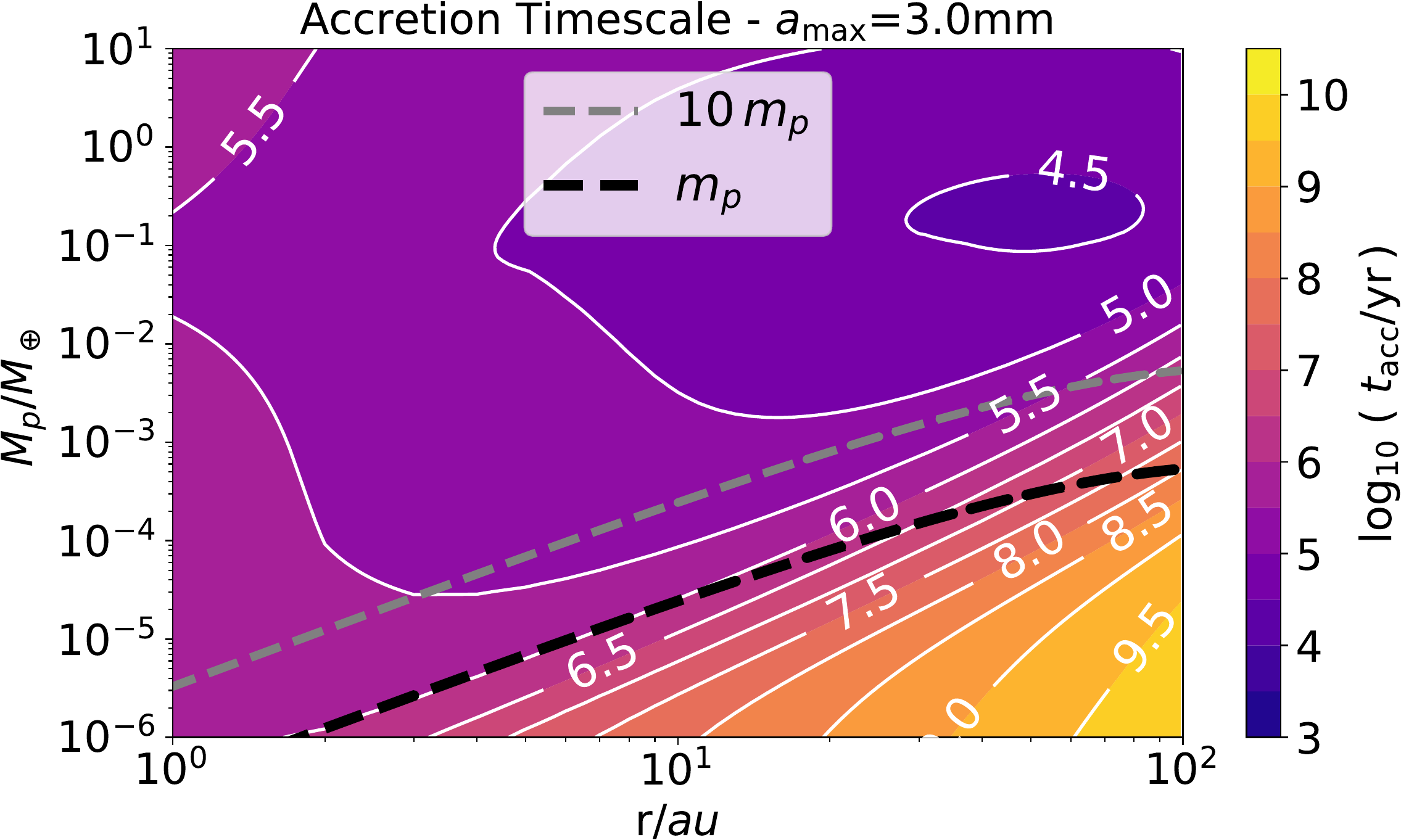}}
    \resizebox{.33\textwidth}{!}{\includegraphics{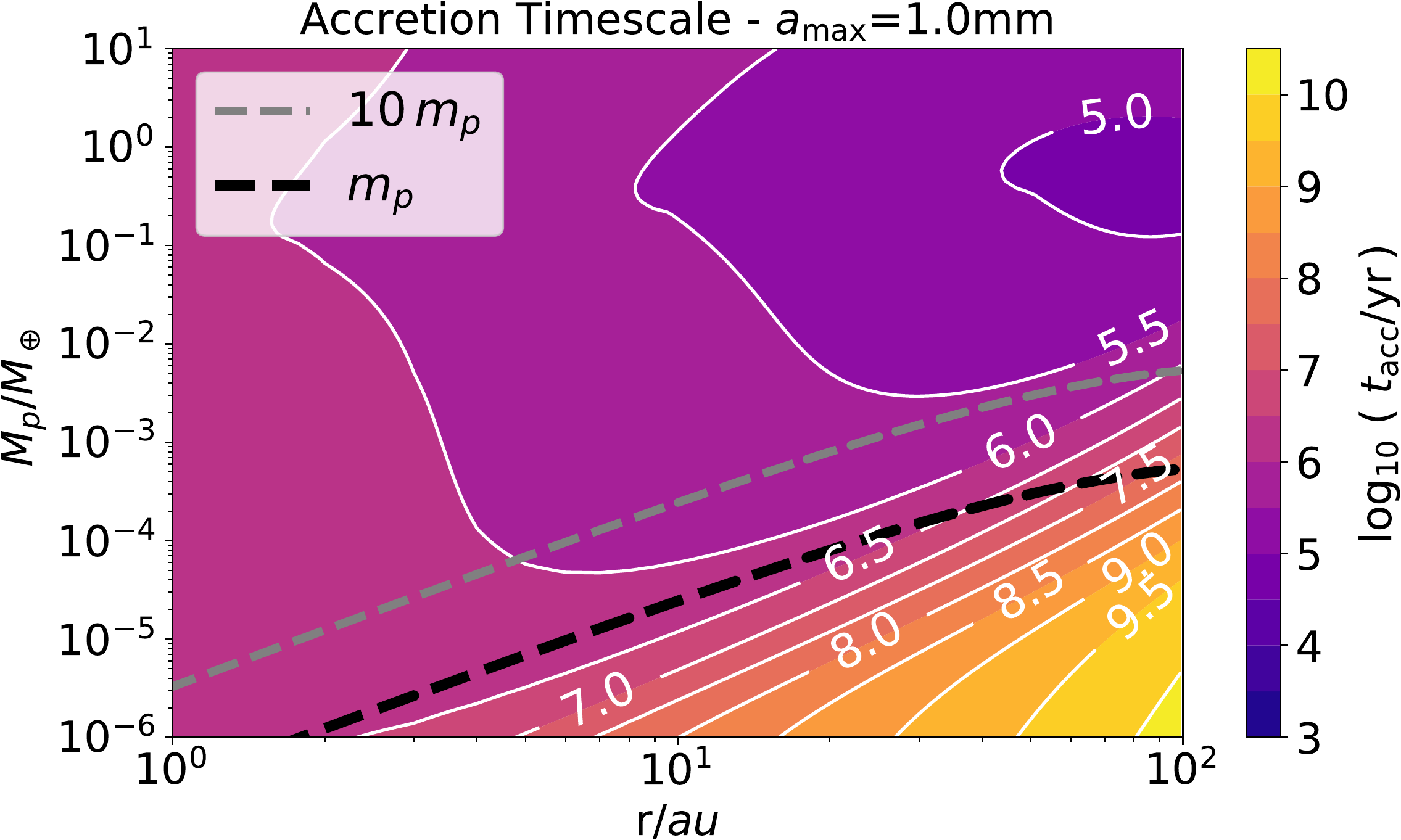}}
    \resizebox{.33\textwidth}{!}{\includegraphics{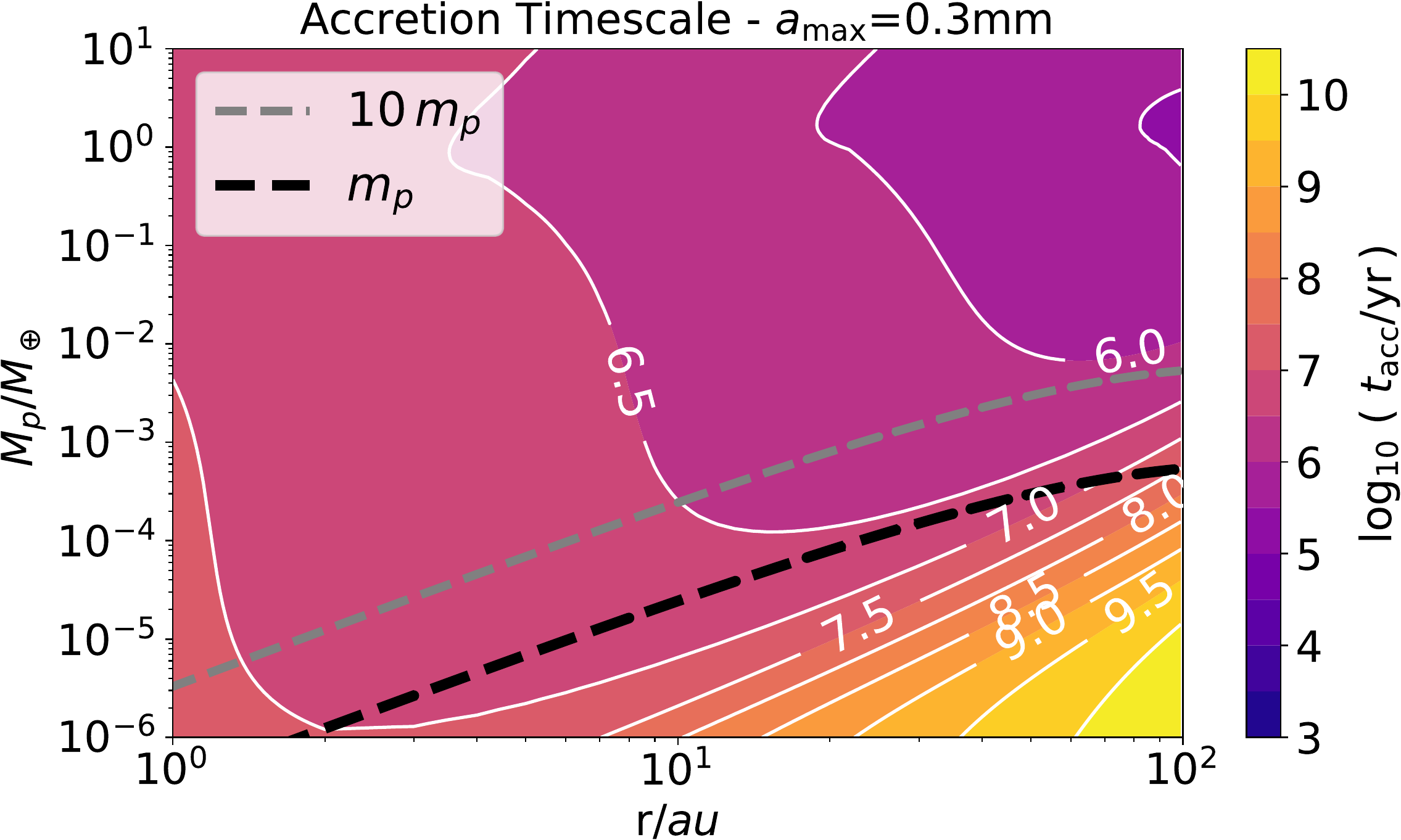}}
  \end{center}
\caption{Same as \fig{fig:IntPebbleAccretion_vs_GrainRadius_distance},
  but exploring the parameter space of maximum grain radius $a_{\rm
    max}$, from left to right: 3\,mm, 1\,mm, and 0.3\,mm. Upper plots
  show mass accretion rate, lower plots the accretion times. The trend
  seen is that Bondi accretion rates decrease with $a_{\rm
    max}$ for the same seed mass and distance. The contour of 6.5 (3Myr) marks the boundary
      where accretion during the lifetime of the nebula is 
      feasible by pebble accretion, without the need for planetesimal
      accretion. This translates into $\approx$ 3\,AU for 100\, km seeds ($10^{-6}
      M_\oplus$), 10\,AU for 0.01 Pluto mass ($2\times 10^{-5}
      M_\oplus$), and up to 30\,AU for 0.1 Pluto mass ($2\times 10^{-4}
      M_\oplus$), for the first two models. The typical products
        of streaming instability grow in Myr timescales except for the
      last model, of maximum grain size 0.3\,mm.}
\label{fig:AccretionTime_vs_GrainRadius_distance}
\end{figure*}

\begin{figure}
  \begin{center}
    \resizebox{\columnwidth}{!}{\includegraphics{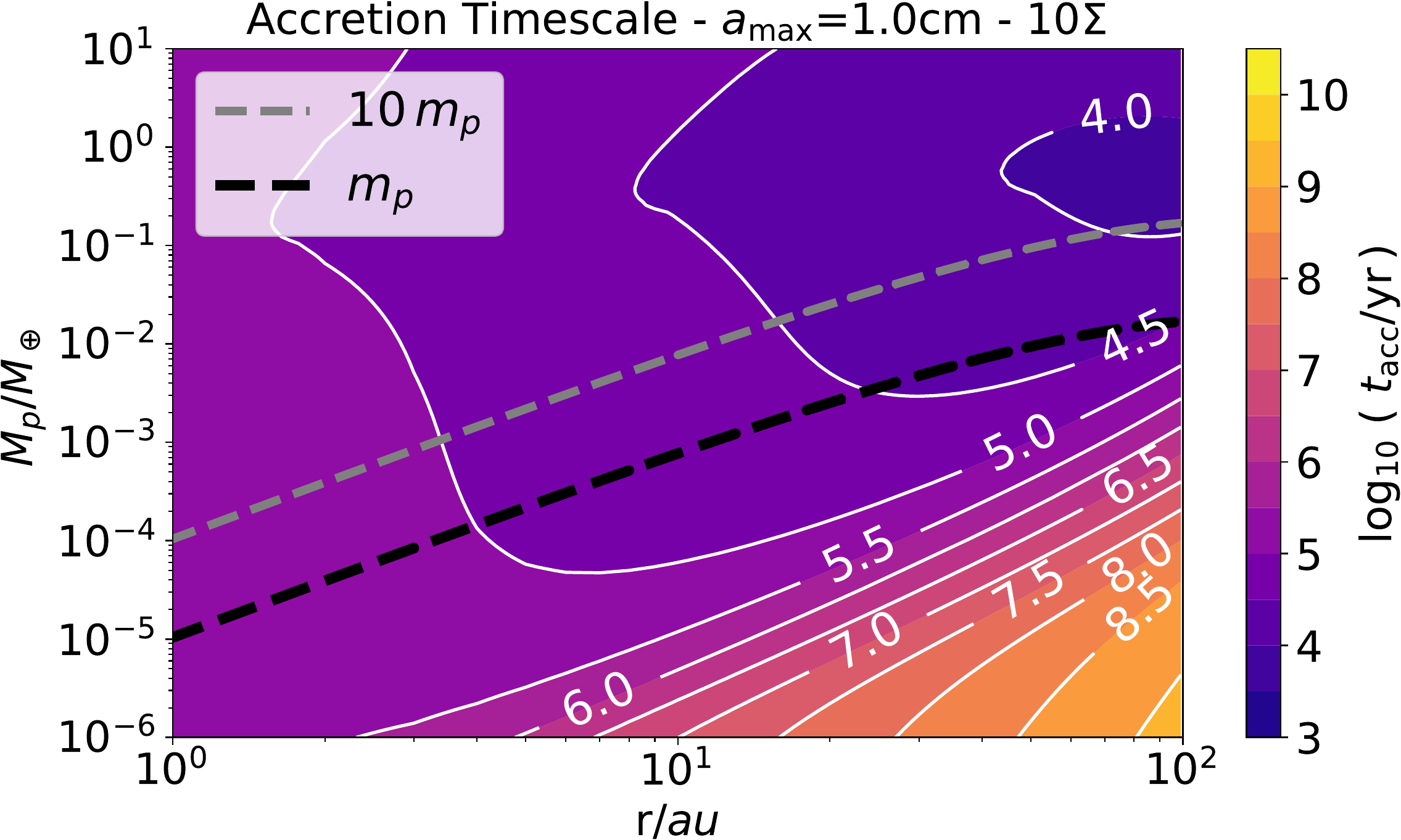}}
  \end{center}
\caption{Same as the right panel of
  \fig{fig:IntPebbleAccretion_vs_GrainRadius_distance}, but for
  10 times the disk mass. Although the Stokes number decreases for the
  same particle radius, the increase in dust mass is the dominant
  effect, and accretion times decrease for the same seed mass and
  distance. Compared to the lower-mass model, the line of 3\,Myr e-folding growth time
  is pushed to about twice the distance, allowing for pebble accretion on
  top of 100\,km seeds ($10^{-6} M_\oplus$) up to 7\,AU. 200\, km
  objects ($10^{-5} M_\oplus$) can
  accrete pebbles efficiently up to 30\,AU. At 40\,AU accretion on
  100\,km seeds takes over 100 Myr and they should remain
  planetesimals, consistent with evidence from the Solar System.}
\label{fig:AccretionTime_10MMSN}
\end{figure}

\begin{figure}
  \begin{center}
    \resizebox{\columnwidth}{!}{\includegraphics{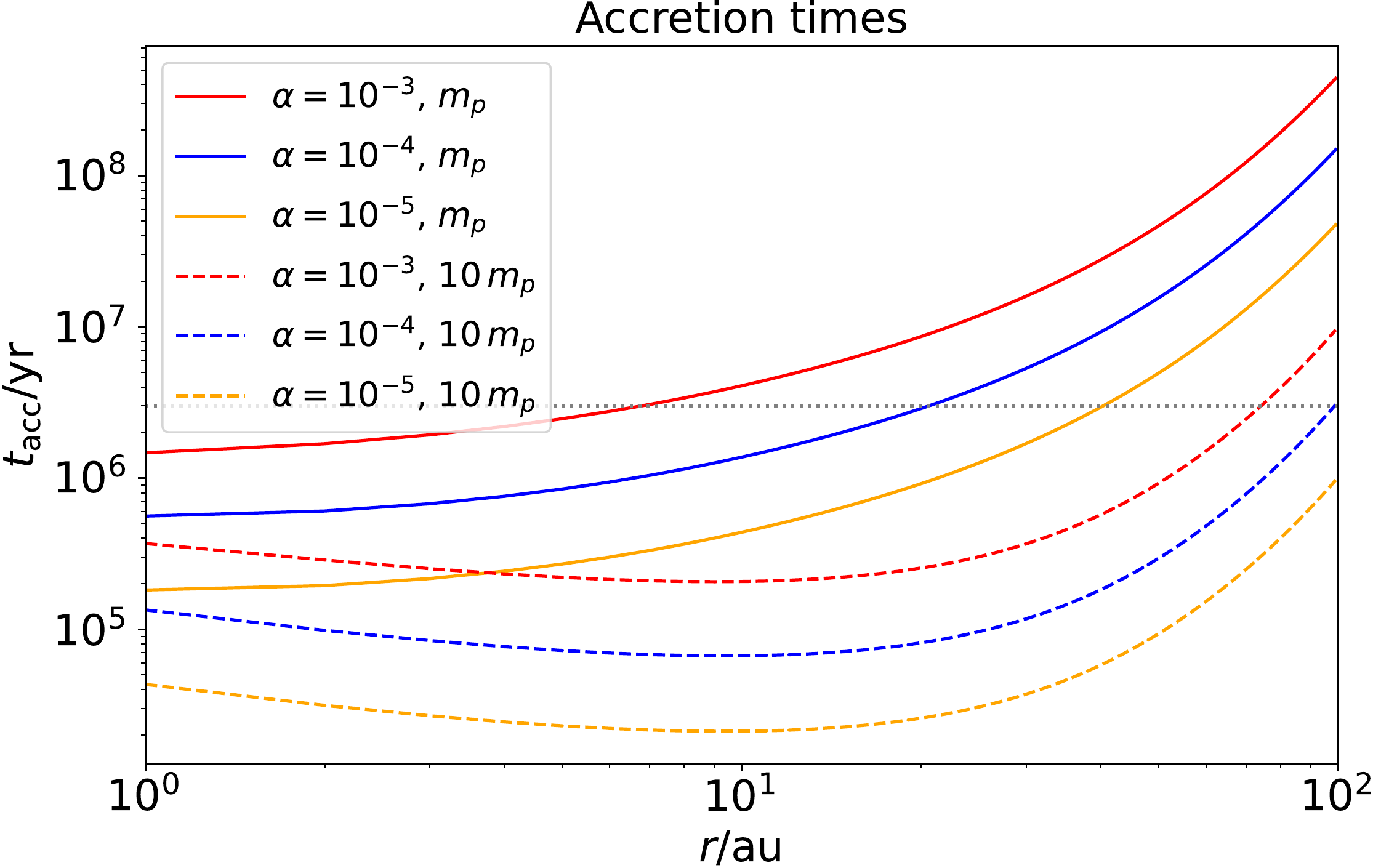}}
  \end{center}
\caption{Polydisperse pebble accretion timescales for different $\alpha$ values for the
    typical masses produced by streaming instability (solid lines),
    and ten times this mass (dashed lines), taken as proxy for the
    end of the streaming instability mass function. The grey dotted line marks
    3\,Myr. For $\alpha=10^{-3}$, the typical seeds only grow within the
    lifetime of the nebula in the
    inner solar system, up to $\approx$5-10\,AU. For lower turbulence,
    $\alpha=10^{-5}$, as most pebbles are sedimented, the distance
    increases to 40\,AU.}
\label{fig:alphatacc}
\end{figure}

We explore now the parameter space of stellocentric distance; the results are shown in
  \fig{fig:PebbleAccretion_vs_GrainRadius_distance}, showing the
  accretion rates at 10, 25, and 40\,AU (notice also we
    decreased $a_{\rm max}$ to 1\,cm). The left plots
  show the integrated mass accretion rates $\dot{M}$, the middle plots
  the distribution $\partial_{\,\ln\,a}\,\dot{M}$, and the right plots the
  distribution normalized by the accretion rate for $a_{\rm max}$. The Hill
  accretion rate decreases only slightly with distance for this model, because the drop in $\varOmega$ and $\varSigma_d$ with
  distance is equally compensated by the increase in the Hill radius.

As for the Bondi regime, we see that at the grain size where
  monodisperse would transition to loosely coupled, polydisperse is
  still about two orders of magnitude more efficient, over all distances
  considered. The seed mass for onset of pebble accretion is also
  pushed down 1 order of magnitude, from $\sim 5\times 10^{-5}$ to
  $\sim 5\times 10^{-6} M_\oplus$ at 10\,AU. This is about
  100-200\,km radius (for internal densities 3.5 and
    0.5 g/cm$^3$, respectively), 
  reaching the range where pebble accretion onto the direct
  products of streaming instability is possible. At 40\,AU the onset
  of pebble accretion is pushed from $\sim 10^{-3} M_\oplus$ in
  monodisperse to $\sim 10^{-4} M_\oplus$
  in polydisperse. A significant reduction, but still
    in the mass range of planetary embryos, so planetesimals
  formed at that distance should remain planetesimals. This is in accordance
  to the solar system constrain given by the existence of the cold classical
  Kuiper Belt objects at 40-50\,AU, presumably undisturbed
  planetesimals.
  
As distance increases, both the
  accretion rate and the size of the best accreted pebble
  decreases. While at 10\,AU the best accreted size for a $10^{-5}
  M_\oplus$ seed (150-300 km radius) is
1\,mm, at 40\,AU it decreases to 10 $\mu$m. This has implications
for the densities of formed objects if the smaller pebbles have
different composition, e.g. the smaller ones being silicate in nature
and the larger ones being icy. Then a planetesimal seed will
preferentially accrete pebbles of rocky composition until it
grows enough in mass to start accreting ices efficiently.

The left panel of \fig{fig:IntPebbleAccretion_vs_GrainRadius_distance} shows the integrated polydisperse pebble accretion
rate as a function of distance, from 1 to 100\,AU. The mass accretion rate of a $10^{-4} M_\oplus$ seed at 20\,AU is about $10^{-10}
M_\oplus  {\rm yr}^{-1}$. The thick black dashed line shows the
  typical mass of objects formed by streaming instability
  \citep{Liu+20,LorekJohansen22}. The thick grey dashed line
    shows 10 times that mass, proxy for the most massive objects
    formed directly by streaming instability.

In the right panel we show the accretion time

   \beq
   t_{\rm acc} \equiv \frac{M_p}{\dot{M}},
   \label{eq:tacc}
   \eeq

   along with the same curves for objects formed by streaming
  instability. The plot shows that a 0.1 Pluto mass ($2\times
  10^{-4} M_\oplus$) seed has e-folding growth time of
  1\,Myr at 20\,AU, and 10\,Myr at 30\,AU; that is, a
  Charon-mass planetary embryo can efficiently increase its mass by Bondi accretion
  during the lifetime of the disk. This implies that the formation of Pluto
  in the solar Nebula as far as 30\,AU is possible by
  Bondi accretion of 10-100 $\mu$m grains onto a 0.1 Pluto mass seed.

  The plot also shows that up to 20\,AU, the objects typically
  formed by streaming instability (thick black dashed line) have
  growth times up to 3\,Myr, within the lifetime of the nebula. Notice that, in the inner
  solar system, Bondi accretion on $10^{-6} M_\oplus$
  seeds ($\approx$ 100\,km radius) at 3\,Myr timescale is possible up to
  3\,AU. We conclude that Bondi accretion directly on planetesimals 
  is possible in the inner solar system, dismissing the need
  for mutual planetesimal collisions as a major contribution to planetary growth.

\subsubsection{Effect of maximum grain size}

In \fig{fig:AccretionTime_vs_GrainRadius_distance} we show the model for 3 different maximum grain sizes,
from left to right: 3\,mm, 1\,mm, and 0.3\,mm. The main feature is
that, as the maximum grain size decreases, the mass accretion rate (accretion time)
for given seed mass at a given distance decreases (increases).

The 3\,Myr contour reaches $10^{-6} M_\oplus$ at 3\,AU for
  $a_{\rm max}$ = 3\,mm, $10^{-6} M_\oplus$ at 2\,AU for 1\,mm, and $10^{-4} M_\oplus$ at 10\,AU
for 0.3mm. The conclusion is similar: Myr-timescale
Bondi accretion on top of 100\,km seeds ($10^{-6}\,M_\oplus$) is
possible in the inner solar system. Except for the model
  with $a_{\rm max}=0.3$\,mm, the typical products of streaming
  instability can grow by pebble accretion in 3\,Myr timescales.

We calculate also a 10$\times$ more massive model. The higher dust mass also
  comes with a higher gas mass, and thus a reduction in Stokes number
  for the same pebble size. It is unclear a priori which effect
  dominates. In \fig{fig:AccretionTime_10MMSN} we show the formation
  times for the model, using $a_{\rm max}=1\,$cm. The formation times
  are overall shorter compared to the right panel of
  \fig{fig:IntPebbleAccretion_vs_GrainRadius_distance}, pushing the
  3\,Myr e-folding contour to double the distance vis-\`a-vis
  the lower mass model (7\,AU for 100\,km, 30\,AU for
  $10^{-2}$ Pluto mass, and 60\,AU for $10^{-1}$ Pluto mass). Even at
  this higher mass model, a 100\,km seed has an e-folding growth time
  of over 100\,Myr at 40\,AU, and should remain planetesimals, as expected.

\subsubsection{Effect of sedimentation}

In \fig{fig:alphatacc} we show the e-folding growth times for
  the planetary seeds formed by streaming instability (typical objects
  and most massive objects), as a function of the turbulent viscosity
  parameter $\alpha$. Its function in the model is only on how it
  influences sedimentation. The grey dotted line in the plot marks
  the threshold of 3\,Myr. For
  moderately high turbulence ($\alpha=10^{-3}$), the typical seeds
  have longer growth times than 3\,Myr already beyond 6\,AU. For lower turbulence,
    $\alpha=10^{-5}$, as most pebbles are sedimented, the distance
    where growth occurs within 3\,Myr increases to 40\,AU. The most
    massive objects, well into the Bondi regime, all have fast growth
    times.
  
\section{Analytical solutions}
\label{sect:analytical}

In this section we derive the analytical solutions in the
  relevant limits of 2D Hill accretion and 3D Bondi accretion. In a
  polydisperse distribution, the pebble scale height is a function
  of pebble radius, so the pebbles are not necessarily all in the 2D
  regime or all in the 3D regime. Also, because the transitions between loosely coupled and
  Bondi, and from Bondi to Hill are St-dependent, the pebbles are
  not all in the same regime of accretion either.

  Yet, in practice these limits still yield reasonably accurate accretion rates. Because the
  distribution is top heavy, the 2D Hill regime is applicable for
  large seed masses, that are accreting in this regime the biggest
  pebbles, which are responsible for most of the mass accretion
  rate. The 3D Bondi regime is applicable as long as $R_{\rm acc} <
  2H_d$ \eqp{eq:xi}, which solving for mass yields

\beq
M_p < \frac{\Delta v \varOmega H_g^2 \alpha}{G \St \left(\St+\alpha\right)} .
\eeq

\noindent Normalizing by the transition mass $M_t$, we find

\beq
\frac{M_p}{M_t} \lesssim \frac{\alpha}{h^2 \St \left(\St+\alpha\right)},
\eeq

\noindent where $h\equiv H_g/r$ is the disk aspect ratio. For $\alpha\sim 10^{-4}$ and $h\sim 10^{-2}$, 3D Bondi accretion should apply close to the
  transition mass, except for big enough pebbles, as expected, because
  these are too sedimented. Yet, as we have established, these pebbles contribute poorly to the
  mass accretion rate. For particles of $\tau_f = t_B$, and assuming $\St \gg \alpha$, we find 

\beq
\frac{M_p}{M_t} \lesssim \left(\frac{\alpha}{h^2}\right)^{1/3},
\eeq

\noindent i.e., within the expected ranges of $\alpha$ and $h$,
  the seed mass for which $\tau_f = t_B$ is within a factor of order
unity from the transition mass. We conclude that a 3D approximation for the
  Bondi regime should lead to acceptable results.
  
We work now the analytical expressions in these limits.

\subsection{Analytical Polydisperse 2D Hill accretion}

We can integrate the polydisperse Hill regime analytically in the 2D
limit by generalizing \eq{eq:monodisperse-2d} with $\varSigma_d$ given by \eq{eq:columndensity} 

\beq
\dot{M}_{\rm 2D, Hill}  =  2\times 10^{2/3} \varOmega R_H^2 \int_0^{a_{\rm max}} \St(a)^{2/3} \, m(a) \, W(a) \, da. 
\label{eq:m2dhillanalytical}
\eeq

\noindent Given the scalings $\St \propto a^{1-q}$, $m\propto
a^{3-q}$, and $W \propto a^{-k}$, the dependency of the
  integrand of \eq{eq:m2dhillanalytical} on $a$ is 

\beq
\frac{\partial \dot{M}(a)}{\partial a }_{\rm 2D, Hill}  \propto  a^{(11-5q -3k)/3}
\eeq

\noindent Integrating it in $a$, we find the exact solution

\beq
\dot{M}_{\rm 2D, Hill} =  \frac{6(1-p)}{14-5q -3k}
\left(\frac{\St_{\rm max}}{0.1}\right)^{2/3} \varOmega \, R_H^2 \, Z\,
\varSigma_g.
\label{eq:2dhill-poly-analytical}
\eeq

\noindent \eq{eq:2dhill-poly-analytical} differs from the monodisperse case
\eqp{eq:monodisperse-2d} by an efficiency factor 

\beq
\left(\frac{\dot{M}_{\rm poly}}{\dot{M}_{\rm mono}}\right)_{\rm 2D, Hill} =\frac{3(1-p)}{14-5q -3k} \left(\frac{\St_{\rm max}}{\St}\right)^{2/3}.
\eeq

\noindent For MRN $(k=3.5)$, $q=0$, and $\St=\St_{\rm max}$, this yields 

\beq
\left(\frac{\dot{M}_{\rm poly}}{\dot{M}_{\rm mono}}\right)_{\rm 2D,
  Hill} ^{k=3.5, q=0} = \frac{3}{7}, 
\eeq

\noindent that is, about 43\% of the monodisperse. Deviations from
this number are due to not all pebbles being in the 2D Hill
regime. For large enough seed mass, the deviations should be small, as
indeed it is seen in the plots of Figs. 3 and 5.

\subsection{Analytical Polydisperse 3D Bondi accretion} 

The Bondi accretion in the 3D regime limit is found by generalizing \eq{eq:monodisperse-3d} with $\rho_{d0}$ given by \eq{eq:rhod0-midplane}

\beqn
\dot{M}_{\rm 3D, Bondi} &=&\frac{4\pi R_B\Delta v^2}{\varOmega} \times \nonumber \\
&&  \int_{0}^{a_{\rm max}} \St \ e^{-2\psi} m(a) f(a) \left[ 1 + 2\left(\St\frac{\varOmega R_B}{\Delta v}\right)^{1/2} e^{-\psi} \right]da,\nonumber\\
\label{eq:3dbondipolyint}
\eeqn

\noindent where we use the shorthand notation

  \beq
  \psi \equiv \chi[\St/(\varOmega t_p)]^\gamma.
  \label{eq:psi}
  \eeq

\noindent We will split \eq{eq:3dbondipolyint} into two integrals

\beqn
\dot{M}_{\rm 3D, Bondi} &=&\frac{4\pi R_B\Delta v^2}{\varOmega}\left[\int_{0}^{a_{\rm max}}  e^{-2\psi} \ \St \ m(a) \ f(a) \ da \right.\nonumber\\
  &+&\left.2 \left(\frac{\varOmega R_B}{\Delta v}\right)^{1/2} \int_{0}^{a_{\rm max}} e^{-3\psi} \ \St^{3/2} \ m(a) \ f(a)  \ da \right].\nonumber\\
\eeqn

\noindent The function $f(a)$ has a dependency on $\sqrt{1+\St/\alpha}$, which
makes these functions non-integrable sauf specific cases. We will thus use the following
approximation, valid at $x\rightarrow 0$ and $x\rightarrow \infty$

\beq
\sqrt{1+x} \approx 1 + \sqrt{x}.
\label{eq:funky-approximation}
\eeq

\begin{figure}
  \begin{center}
    \resizebox{\columnwidth}{!}{\includegraphics{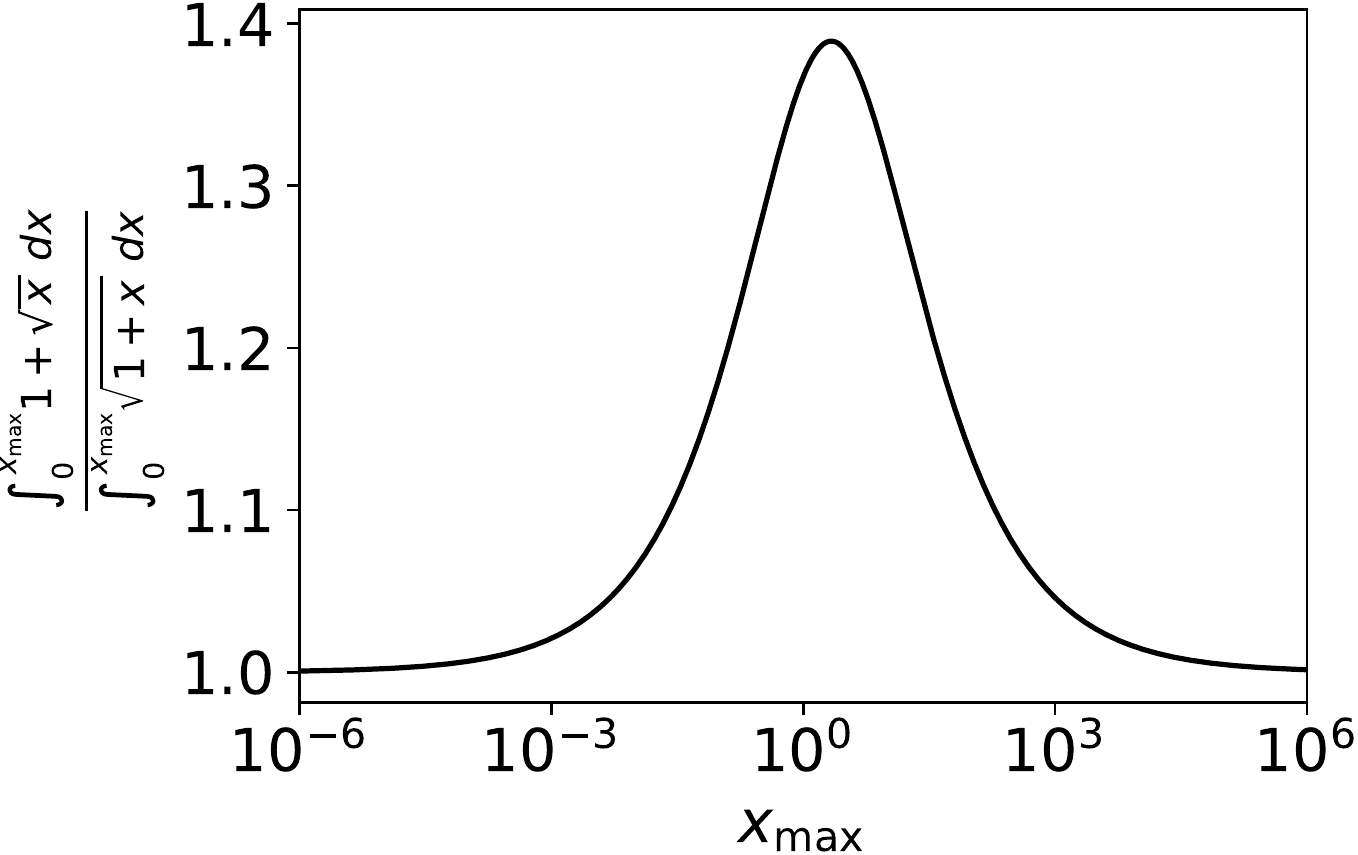}}
\end{center}
\caption{Approximating $\sqrt{1+x}$ in by $1+\sqrt{x}$ (the
  asymptotic expansion for $x\rightarrow 0$ and $x\rightarrow \infty$), to make the sedimented
  midplane distribution integrable analytically. As long as the function is
  integrated to a large value of $x_{\rm max}$, the error incurred is small.}
\label{fig:funky-approximation-works}
\end{figure}

\noindent While the error incurred with this approximation at $x\approx 1$
  can be large, we are interested in the definite integral from 0 to $x_{\rm max}$. In this case, the error decreases if
  the range of integration is large enough, tending to zero for $x_{\rm
    max} \rightarrow \infty$, as shown in
  \fig{fig:funky-approximation-works}. Confident in the accuracy of
  \eq{eq:funky-approximation}, we write the approximate solution 

\beqn
\dot{M}_{\rm 3D, Bondi} &\approx&\frac{3 (1-p)Z\varSigma_g R_B\Delta v^2}{\sqrt{2\pi} H_g  \varOmega\rho_{\bullet}^{(0)} a_{\rm max}^{4-k}}\times\nonumber\\
&&\left[ \int_{0}^{a_{\rm max}} e^{-2\psi} \ \St  \ m(a) \ a^{-k} da\right.\nonumber\\
&+&\alpha^{-1/2}\int_{0}^{a_{\rm max}} e^{-2\psi} \ \St^{3/2}  \ m(a) \ a^{-k} da \nonumber\\
  &+&2 \left(\frac{\varOmega R_B}{\Delta v}\right)^{1/2} \int_{0}^{a_{\rm max}} e^{-3\psi} \ \St^{3/2} \ m(a) \ a^{-k}  da \nonumber\\
  &+&\left.2 \left(\frac{\varOmega R_B}{\alpha\Delta v}\right)^{1/2} \int_{0}^{a_{\rm max}} e^{-3\psi} \ \St^{2} \ m(a) \ a^{-k}  da \right].
\label{eq:3dbondi-4integrals}
  \eeqn

The four integrals are of the form below, for which there is an
analytical solution in terms of lower incomplete gamma functions

\beq
\int_0^{a_{\rm max}} e^{-j a^s} a^b da =\frac{\gamma_l\left(\frac{b+1}{s}, ja_{\rm max}^s\right)}{sj^{(b+1)/s}}.
\eeq

We thus write the solution of \eq{eq:3dbondi-4integrals}

\beqn
\dot{M}_{\rm 3D, Bondi} &\approx&C_1\frac{\gamma_l\left(\frac{b_1+1}{s}, j_1 a_{\rm max}^{s}\right)}{sj_1^{(b_1+1)/s}} +C_2\frac{\gamma_l\left(\frac{b_2+1}{s}, j_2 a_{\rm max}^{s}\right)}{sj_2^{(b_2+1)/s}} +\nonumber\\
&&C_3\frac{\gamma_l\left(\frac{b_3+1}{s}, j_3 a_{\rm max}^{s}\right)}{sj_3^{(b_3+1)/s}} +C_4\frac{\gamma_l\left(\frac{b_4+1}{s}, j_4 a_{\rm max}^{s}\right)}{sj_4^{(b_4+1)/s}} ,
\label{eq:3dbondi-poly-analytical}
\eeqn

\noindent where the coefficients are

\beqn
s&=&\gamma(1-q)\\
b_1&=&4-2q-k\\
b_2=b_3&=&(9-5q-2k)/2\\
b_4 &=& 5-3q-k\\
\St^\prime &=&\frac{\pi}{2\varSigma_g}\rho_\bullet^{(0)}a_{\rm max}^{q}\\
j^\prime &=&\chi\left(\frac{\St^\prime}{\varOmega t_p}\right)^\gamma \\
j_1=j_2&=&2j^\prime\\
j_3=j_4&=&3j^\prime\\
m^\prime &=&  \frac{4\pi}{3}\rho_\bullet^{(0)}a_{\rm max}^{q}\\
K &=&\frac{3 (1-p)Z\varSigma_g R_B\Delta v^2}{\sqrt{2\pi} H_g  \varOmega\rho_{\bullet}^{(0)} a_{\rm max}^{4-k}} \\
C_1&=&K \St^\prime m^\prime\\
C_2&=&K \St^{\prime 3/2} m^\prime\alpha^{-1/2}\\
C_3&=&2K \St^{\prime 3/2} m^\prime \left(\frac{\varOmega R_B}{\Delta v}\right)^{1/2} \\
C_4&=&2K \St^{\prime 2}     m^\prime \left(\frac{\varOmega R_B}{\alpha\Delta v}\right)^{1/2}
\eeqn

\begin{figure}
  \begin{center}
    \resizebox{\columnwidth}{!}{\includegraphics{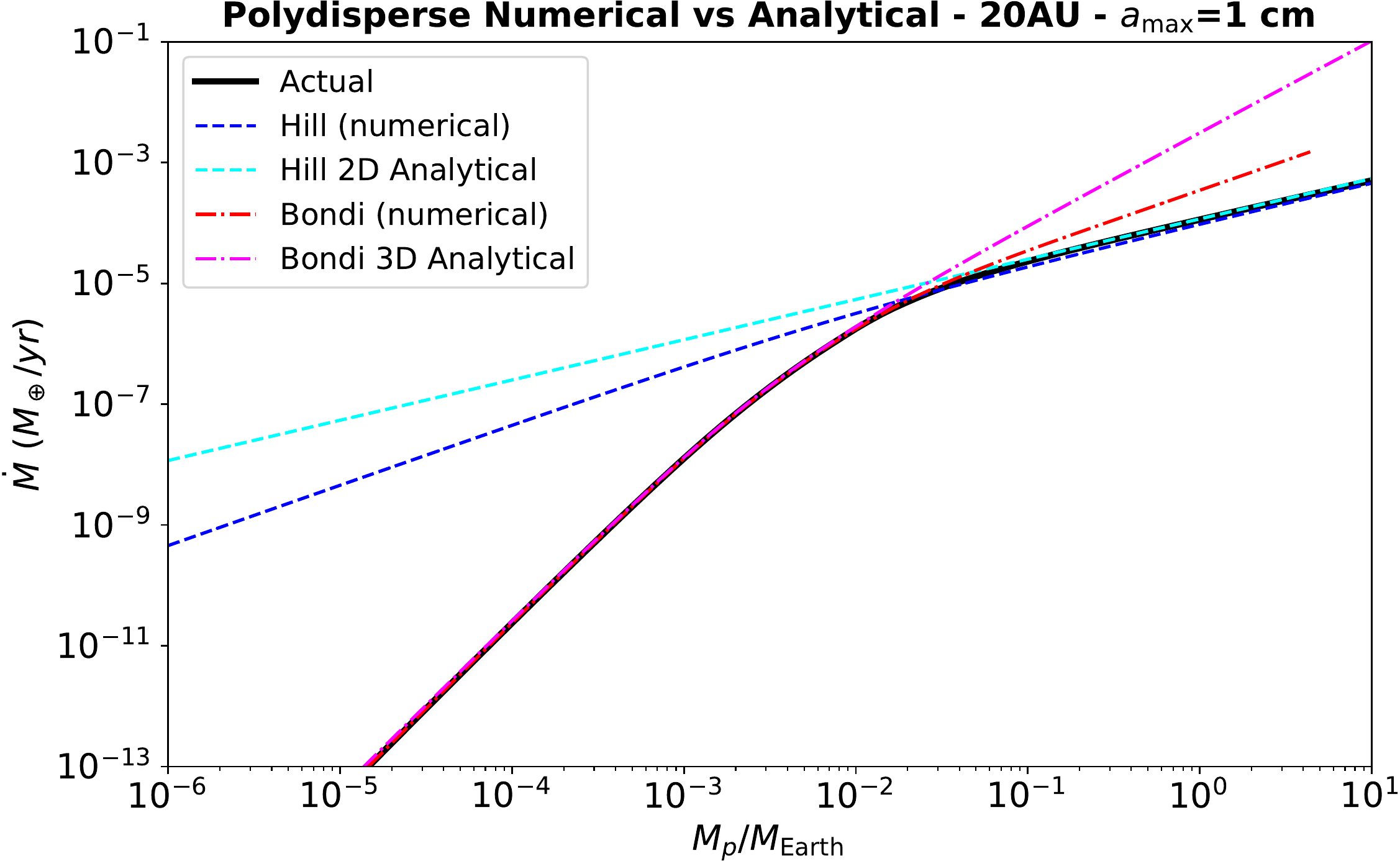}}
\end{center}
\caption{Agreement between the numerically calculated
    polydisperse pebble accretion rate and the analytical solutions
    for 2D Hill accretion \eq{eq:2dhill-poly-analytical} and 3D Bondi
    accretion \eq{eq:3dbondi-poly-analytical}. While the Hill solution
    is exact, the Bondi solution is approximate. Yet, the agreement seen
    is excellent, because the best accreted pebbles in this regime are
    in the 3D range.}
\label{fig:NumericalVsAnalytical}
\end{figure}

\fig{fig:NumericalVsAnalytical} shows that the agreement between
the numerical integration of \eq{eq:mass-accretion-rate-integrand} and the analytical solutions
(\eq{eq:2dhill-poly-analytical} and \eq{eq:3dbondi-poly-analytical}) is excellent
in the range of validity. Having \eq{eq:2dhill-poly-analytical} and
\eq{eq:3dbondi-poly-analytical} as analytical
expressions is of great interest for future studies including pebble
accretion analytically, instead of having to integrate the
mass accretion rates numerically with the particle size
distributions.

\begin{figure}
  \begin{center}
    \resizebox{\columnwidth}{!}{\includegraphics{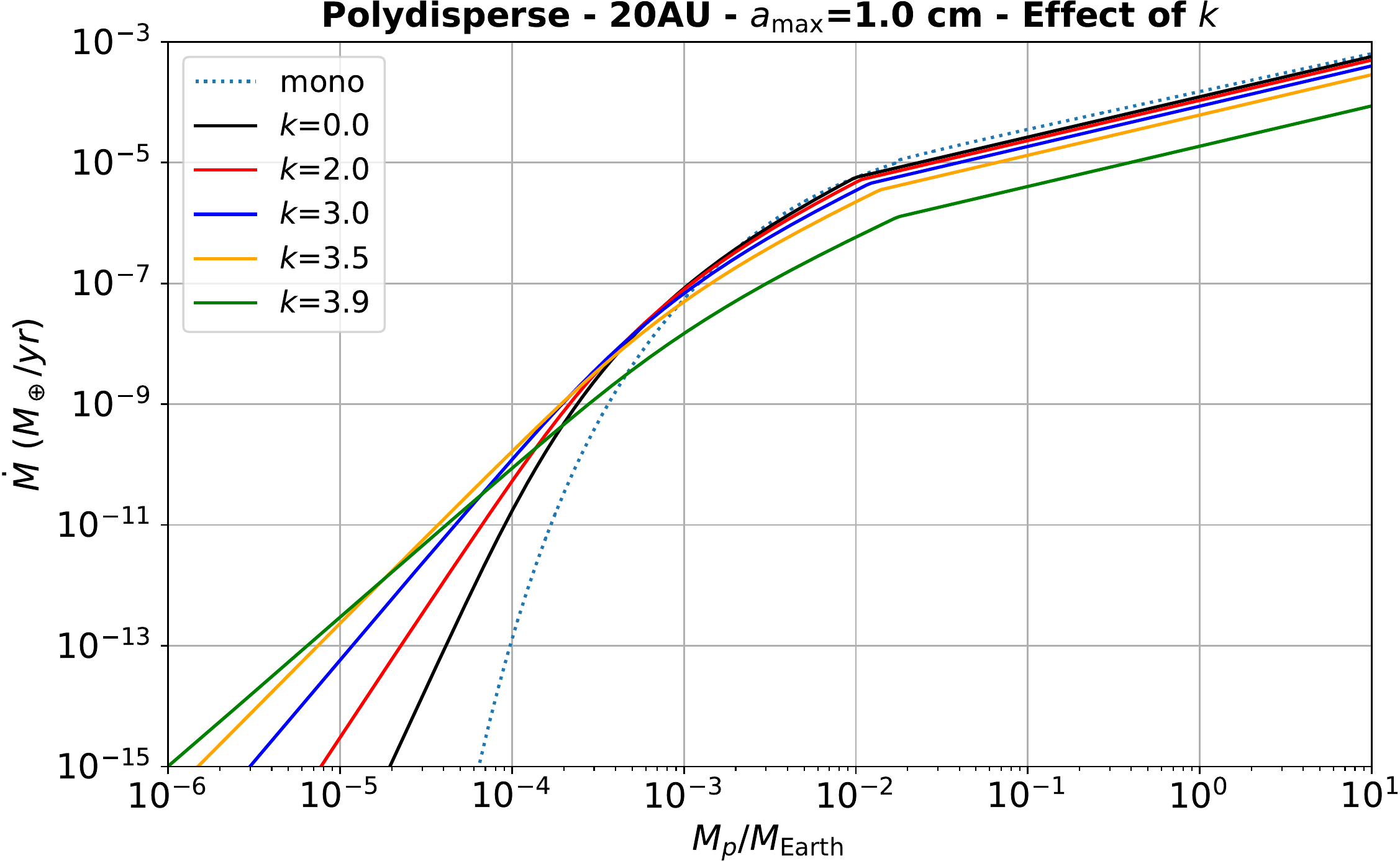}}
\end{center}
\caption{Effect of varying the slope $k$ of the grain size
    distribution. The Hill regime is relatively insensitive to $k$, as
    this regime is dominated by the largest grains. For Bondi
    accretion, the mass accretion rates increase significantly as the
    slope steepens.}
\label{fig:keffect}
\end{figure}

\section{Effect of slope $k$ of grain size distribution}

So far we have considered only the MRN value for
  the index $k$ of the grain size distribution
  \citep{Mathis+77}, but this index should depend on the collisional
  evolution, velocities, and material strength of the pebbles
\citep{KobayashiTanaka10,Kobayashi+16}. Having found the analytical
solution for the accretion rates, we can more easily determine the
impact of varying this parameter, which we show in \fig{fig:keffect}.
As the slope steepens, the mass accretion rate decreases in the Hill regime. Compared to
monodisperse, the effect is small, but accelerates as $k$
approaches 4. This insensitivity is expected, as the Hill regime is
dominated by the larger grains. The effect on mass accretion rate
is more pronounced for the Bondi regime, as expected, as the amount
of mass in different grain sizes more strongly affects this accretion
regime. As the slope steepens and more mass is made available in small
grain sizes, the accretion rates onto smaller planetesimal seeds
increases, although the effect is nonlinear.

\section{Limitations}

We are limited in this work by the vast expanses of the parameter
space and by the circular restricted 3-body problem
solution that forms for the underlying assumption of the gas and
pebble flow. While the former would be a valiant endeavour, it is not the scope of this
work to derive results applicable to all
possible situations, but to derive the model in first place. As such, we kept our
equations general in metallicity, internal density, and grain size
distribution, but apply it mostly for $Z=0.01$, $\rho_\bullet^{(0)} =
3.5$\,g\,cm$^{-3}$, $q=0$. These parameters will vary with dust
drift (lower the metallicity), composition (varying the internal
density if ices or silicates), and porosity.

As for going beyond the circular restricted 3-body problem, recently the impact of the gravity of the
planetary seed on the accretion flow has been calculated
from hydrodynamical simulations \citep{OkamuraKobayashi21}, for the Hill regime \citep{KuwaharaKurokawa20a} and for the Bondi regime
\citep{KuwaharaKurokawa20b}. In the Bondi regime, the trajectories are
modified for $\St \lesssim 10^{-3}$, with the gas flow reducing the
accretion rate. Thus, \eq{eq:rhatbondi} is overestimated for small
$\St$. We find that the best accreted pebbles, that give the bulk of
the boost in Bondi accretion, are slightly above the $\St \sim 10^{-3}$ transition found by
\citep{KuwaharaKurokawa20b}; as such, this aspect of our results are not severely
affected by the planet-induced flow.

\section{Conclusion}
\label{sect:conclusion}

In this paper, we worked out the theory of polydisperse pebble
accretion, finding analytical solutions when possible. Our
  main findings are as follows:

\begin{itemize}

\item We find that polydisperse Bondi accretion is 1-2 orders of
  magnitudes more efficient than in the monodisperse case, This is
because the best-accreted pebbles in the Bondi regime are those of
friction time similar to Bondi time, not the largest pebbles present.
 The large pebbles, although dominating the mass budget, 
are weakly coupled across the Bondi radius and thus accrete
poorly. The pebbles that are optimal for Bondi accretion may
contribute less to the mass budget, but their enhanced accretion
significantly impacts the mass accretion rate. 

\item The onset of polydisperse pebble accretion is extended by 1-2
  orders of magnitude lower in mass compared to monodisperse, for the
  same reason.   The onset of pebble accretion with Myr-timescales
  reaches 100-350\,km sized objects depending on stellocentric
  distances and disk model. For the model considered, Bondi accretion on Myr timescales, within the lifetime of the disk, is
  possible on top of $10^{-6} M_\oplus$ (100\,km) seeds up to 4\,AU, on top of $10^{-5} M_\oplus$ (200\,km)
  seeds up to 10\,AU, and on $10^{-4} M_\oplus$ (350\,km) seeds up to
  30\,AU. A model 10 times more massive doubles these distances.

  \item In all models considered, at 40\,AU a 100\,km seed has growth
    time over 100\,Myr, and should thus remain as planetesimals, in
    accordance with the existence of the cold classical Kuiper Belt
    population, presumably undisturbed planetesimals.

\item We find the analytical solution of the stratification integral,
  and thus the exact solution for the 3D-2D transition \eqp{eq:monodisperse-general},

\item We find analytical solutions for the polydisperse 2D
  Hill \eqp{eq:2dhill-poly-analytical} and 3D
  Bondi regime \eqp{eq:3dbondi-poly-analytical}. For the MRN
  distribution, the Hill accretion is a factor 3/7 (about 42\%) as efficient in
polydisperse than monodisperse. 

\end{itemize}

The fact that Myr-growth timescales, within the lifetime of the disk, is
possible for polydisperse pebble accretion onto 100-350\,km seeds over a
significant range of the parameter space, has significant
implications. This mass range overlaps with the high mass end of
  the planetesimal initial mass function \citep{Johansen+15,Schafer+17,Li+19}, and thus pebble accretion is
  possible directly following formation by streaming instability,
  removing the need for planetesimal accretion. This conclusion is
  supported by the lack of of craters generated by 1-2\,km
  on Pluto \citep{Singer+19}, and recent findings by
  \cite{LorekJohansen22} that planetesimal accretion are not able to sustain accretion rates
  beyond 5\,AU.

While we do most of our numerical solutions with constant
$\rho_\bullet$, we keep the analytical solutions general for varying
this parameter, expecting that smaller pebbles should be of lower
density, and the bigger pebbles of higher density, reflecting
different compositions \citep{Morales+16}. We notice that as the
distance increases, the pebble size that maximizes pebble accretion
is increasingly smaller. This implies the possibility of a two-mode
formation of Kuiper belt objects: streaming instability of the largest
pebbles forming icy objects of the order of $\gtrsim$ 100\,km in
diameter, followed by pebble accretion leading to objects of the order
of 1000 km, where silicates are incorporated mostly at the pebble
accretion stage, due to their low Stokes number. This scenario would
lead to a different composition for the smaller objects,
mostly formed by ice streaming instability, and the larger objects,
grown by ice and silicate pebble accretion on top of the icy
planetesimal seeds. A continuum of rock-to-ice fraction should be
produced. Indeed a trend is clear in the Kuiper belt, of constant
density around 0.5 g\,cm$^{-3}$ for the smaller objects (diameter less
than 500 km), and increasing for larger objects \citep{Brown12,Grundy+15,McKinnon+17}. We will explore how
our findings in this paper can reproduce this result in a future work.

\acknowledgments

WL acknowledges support from the NASA Theoretical and Computational
Astrophysical Networks (TCAN) via grant 80NSSC21K0497, from the
  NASA Emerging Worlds program via grant 22-EW22-0005, and by NSF via grant
AST-2007422. AJ is supported by the Swedish Research Council (Project
Grant 2018-04867), the Danish National Research Foundation (DNRF Chair
grant DNRF159), and the Knut and Alice Wallenberg Foundation
(Wallenberg Academy Fellow Grant 2017.0287). A.J. further thanks the
European Research Council (ERC Consolidator Grant 724 687-PLANETESYS),
the Göran Gustafsson Foundation for Research in Natural Sciences and
Medicine, and the Wallenberg Foundation (Wallenberg Scholar KAW
2019.0442) for research support. MHC is supported by grant
22-EW22-0005 from the NASA Emerging Worlds program.
We acknowledge conversations with Andrew Youdin, Jake Simon, Orkan
Umurhan, Debanjan Sengupta, and Daniel Carrera.

\bibliographystyle{apj}
%\bibliography{./master.bib}

\input{main.bbl}
\end{document}

%% file: preamble.tex
\usepackage{apjfonts}

\usepackage{graphicx}
\usepackage{color}

\usepackage{amsmath,amssymb}

\usepackage{xfrac}
\usepackage{sidecap}
\usepackage{url}
\usepackage{hyperref}
\usepackage{enumitem}
\newlength{\figspace}
\newcommand{\beq}{\begin{equation}}
\newcommand{\eeq}{\end{equation}}
\newcommand{\beqn}{\begin{eqnarray}}
\newcommand{\eeqn}{\end{eqnarray}}

\newcommand{\St}{\mathrm{St}}

\newcommand{\Stp}{\mathrm{St_p}}

\newcommand{\Eq}[1]{Eq.~(\ref{#1})}
\newcommand{\Eqs}[2]{Eqs.~(\ref{#1})~and~(\ref{#2})}

\newcommand{\eq}[1]{\Eq{#1}}
\newcommand{\eqp}[1]{(Eq.~\ref{#1})}
\newcommand{\eqs}[2]{\Eqs{#1}{#2}}

\newcommand{\Fig}[1]{Fig.~\ref{#1}}

\renewcommand{\fig}[1]{\Fig{#1}}

\renewcommand{\table}[1]{Table~\ref{#1}}
\newcommand{\sect}[1]{Sect.~\ref{#1}}

\definecolor{brown}{rgb}{0.42,0.24,0.07}
\definecolor{darkgreen}{rgb}{0.0,0.6,0.00}
\definecolor{purple}{rgb}{0.7,0.0,0.7}
\definecolor{black}{rgb}{0.0,0.0,0.0}

\def\apj{\rm ApJ}
\def\apjl{\rm ApJL}

\def\mnras{\rm MNRAS}
\def\nat{\rm Nature}

\def\aap{\rm A\&A}

\def\araa{\rm ARA\&A}
\def\icarus{\rm Icarus}